\providecommand{\keptcontentend}{}

\documentclass[conf]{new-aiaa}

\usepackage{xcolor}
\definecolor{GuidanceBlue}{HTML}{005BBB}  
\definecolor{WarningRed}{HTML}{D40000}    

\usepackage[utf8]{inputenc}  
\usepackage{graphicx}
\usepackage{amsmath}
\usepackage[version=4]{mhchem}
\usepackage{siunitx}
\usepackage{fontawesome5}
\usepackage{array}            
\usepackage{booktabs}         
\usepackage{multirow}         
\usepackage{makecell}         
\usepackage{tabularx}         
\usepackage{longtable}        
\usepackage{colortbl}         
\usepackage{xcolor}           
\usepackage[normalem]{ulem}   
\usepackage{subcaption}
\usepackage{rotating}     
\usepackage{pdflscape}    
\usepackage{tabularx}     
\usepackage{longtable}    
\usepackage{adjustbox}    
\usepackage{array}        
\definecolor{GuidanceBlue}{HTML}{005BBB}
\definecolor{RowBlue}{HTML}{E8F1FB}
\definecolor{RowGray}{HTML}{F7F7F7}
\definecolor{LightRed}{HTML}{FDECEC}
\definecolor{LightGreen}{HTML}{E9F7EF}

\renewcommand{\arraystretch}{1.05} 
\title{A Survey of Security Challenges and Solutions for Advanced Air Mobility and eVTOL Aircraft}

\author{Mahyar Ghazanfari, Iman Sharifi, Peng Wei}
\affil{George Washington University, Washington, D.C. 20052, USA}

\author{Noah Dahle, Abel Diaz Gonzalez, Austin Coursey, Bryce Bjorkman, Cailani Lemieux-Mack, Robert Canady, Abenezer Taye, Bryan C.\ Ward, Xenofon Koutsoukos, Gautam Biswas}

\affil{Vanderbilt University, Nashville, TN, 37212, USA}
\author{Maheed H. Ahmed, Hyeong Tae Kim, Mahsa Ghasemi, Vijay Gupta}

\affil{Purdue University, West Lafayette, IN, 47907, USA}
\author{Filippos Fotiadis, Ufuk Topcu}
\affil{The University of Texas at Austin, Austin, TX, 78712, USA}

\author{Junchi Lu, Alfred Chen}
\affil{University of California, Irvine, Irvine, CA, 92697, USA}

\author{Abdul kareem Ras, Nischal Aryal, Amer Ibrahim, Amir Shirkhodaie}
\affil{Tennessee State University, 3500 John Merritt Blvd, Nashville, TN, 37221, USA }

\author{Heber Herencia-Zapana, Saqib Hasan, Isaac Amundson}
\affil{Advanced Research and Technology, Collins Aerospace, Cedar Rapids, IA, 52498, USA}

\begin{document}

\maketitle

\begin{abstract}
This survey reviews the existing and envisioned security vulnerabilities and defense mechanisms relevant to Advanced Air Mobility (AAM) systems, with a focus on electric vertical takeoff and landing (eVTOL) aircraft. Drawing from vulnerabilities in the avionics in commercial aviation and the automated unmanned aerial systems (UAS), the paper presents a taxonomy of attacks, analyzes mitigation strategies, and proposes a secure system architecture tailored to the future AAM ecosystem. The paper also highlights key threat vectors, including Global Positioning System (GPS) jamming/spoofing, ATC radio frequency misuse, attacks on TCAS and ADS-B, possible backdoor via Electronic Flight Bag (EFB), new vulnerabilities introduced by aircraft automation and connectivity, and risks from flight management system (FMS) software, database and cloud services. Finally, this paper describes emerging defense techniques against these attacks, and open technical problems to address toward better defense mechanisms. 
\end{abstract}

\section{Introduction}
\smallskip

\lettrine[lines=3, lhang=0.1, loversize=0.15]{A}{dvanced Air Mobility (AAM)} represents a transformative approach to aviation, aiming to integrate a new class of electric, highly automated aircraft—including eVTOL vehicles—into the national airspace to provide short- to medium-range transportation solutions for both passengers and cargo. AAM addresses the growing need for efficient, scalable mobility across urban and regional settings by leveraging advancements in electric propulsion, autonomy, and coordinatedair traffic management \cite{chancey2023hat, maven2022vertiports}.

Unlike conventional rotorcraft or general aviation systems, AAM platforms feature novel connected aircraft and rely on highly automated operations across all phases of flight. These aircraft are expected to operate in complex, high-density, and dynamically evolving airspace environments that span a broad range of altitudes. Within the AAM ecosystem, Urban Air Mobility (UAM) is a critical subset that focuses on operations within densely populated metropolitan areas, whereas Regional Air Mobility (RAM) extends the operational range by connecting smaller cities and rural communities over longer distances \cite{ram2021}.

While much attention has been directed toward physical infrastructure and regulatory policy, securing the digital infrastructure, spanning navigation, communication, flight control, and system data integrity, is equally critical. As the National Aeronautics and Space Administration (NASA), the Federal Aviation Administration (FAA), and other stakeholders advance the AAM framework, the integration of eVTOL operations into the National Airspace System (NAS) introduces a host of technical and security challenges. Chief among them is cybersecurity, which has emerged as a vital concern. The increasing reliance on GPS navigation, automated flight control, aircraft connectivity, data links, and cloud-connected services renders eVTOL platforms especially vulnerable to cyber threats such as spoofing, jamming, man-in-the-middle attacks, and denial-of-service (DoS) disruptions.

Building and maintaining public trust in AAM operations, particularly those involving eVTOL aircraft, requires more than compliance with airworthiness standards and operational efficiency. It demands resilient architectures, secure communication channels, and end-to-end data protection. As noted by the Secure Airspace Technology Group (SATG), cybersecurity must be a foundational component of AAM system design rather than a reactive consideration \cite{freeman2020cyber}. This paper explores the cybersecurity landscape of eVTOL and AAM platforms by surveying known vulnerabilities of commercial avionics, discussing envisioned risks introduced by aircraft automation and connectivity, analyzing potential attack surfaces, and reviewing recent advancements in defense strategies aimed at ensuring system integrity, availability, and confidentiality.

While prior surveys have examined cybersecurity risks in aviation and UAM more broadly, their focus differs from the scope of this work. Existing reviews either analyze high-level threat taxonomies across conventional aircraft systems \cite{habler2023assessing} or provide generalized vulnerability overviews for early UAM concepts \cite{tang2020review}. In contrast, our work concentrates specifically on eVTOL platforms and their operational ecosystem, offering a deeper, system-oriented analysis of vulnerabilities and defense mechanisms tailored to the sensing, autonomy, communication, and cloud-integration requirements unique to AAM environments.

The remainder of this paper is organized as follows. Section~\ref{sec: Background} provides background on the fundamental concepts of AAM and eVTOL systems. Section~\ref{sec: Existing Vulnerabilities and Defense Mechanisms} outlines current vulnerabilities and potential cyberattacks targeting these platforms, and presents defense strategies and mitigation techniques corresponding to these threats. Finally, section ~\ref{sec: Conclusion} concludes the paper.

\section{Background} \label{sec: Background}

\smallskip

AAM is an evolving aviation paradigm that seeks to integrate next-generation, highly automated, and often electric aircraft into existing airspace systems to provide efficient, safe, and flexible transportation for both passengers and cargo. AAM encompasses a wide spectrum of operations, from short-range urban flights to regional and intercity missions, and aims to alleviate surface transportation congestion, increase connectivity, and improve accessibility in both densely populated and underserved areas. The AAM framework is being actively advanced by the FAA, NASA, and a broad range of industry and academic stakeholders to enable scalable and equitable air mobility solutions across diverse operational environments.

Within this broader framework, UAM is a key operational subset of AAM focused specifically on transportation within and around metropolitan regions. UAM introduces aerial mobility solutions that are optimized for high-density, short-range travel, particularly in congested urban areas. A central enabler of UAM operations is the eVTOL aircraft, which features electric propulsion, vertical lift capability, and reduced noise signatures. These vehicles are designed for point-to-point operations between dedicated infrastructure nodes, such as vertiports and vertistops, often with autonomous or remote piloting capabilities. Their compact footprint and operational flexibility make them well-suited for high-frequency flights in urban airspaces.

A notional UAM architecture, depicted in \cite{faaUAM2023}, illustrates the relationships among the principal actors involved in UAM operations. This architecture operates within the larger AAM ecosystem and is centered around a federated service network that manages information flow between the FAA, UAM operators, infrastructure providers, and public stakeholders. It emphasizes interoperability through standardized data formats and communication protocols, enabling safe, coordinated, and scalable integration of UAM into the NAS.

Communication systems are foundational to the safe and scalable operation of eVTOL aircraft. These systems must accommodate dense, three-dimensional urban traffic, ensure low-latency data exchange, and operate reliably under variable environmental and network conditions. eVTOL communication architectures are centered around air-to-ground, air-to-air, and potentially satellite communication to support command and control (C2), situational awareness, and coordination among vehicles and infrastructure. Communication requirements for eVTOLs include ultra-reliable coverage, high data rate (up to 100 Mbps for remote piloting), and low end-to-end latency (as low as 10 ms for safety-critical maneuvers) \cite{abuzaid2023communications}. Fifth-generation (5G) cellular networks and their upcoming 6G counterparts offer promising infrastructure for wide-area eVTOL connectivity. These technologies support low-latency, high-bandwidth communication and can be supplemented with rooftop-mounted base stations or dedicated urban nodes to improve coverage in dense environments \cite{abuzaid2023communications}.


Navigation systems for eVTOLs combine conventional aviation tools with advanced onboard autonomy to support precise flight in complex urban airspaces. Position estimation is primarily achieved through Global Navigation Satellite System (GNSS) and Inertial Navigation System (INS) integration, in which GPS signals are fused with inertial measurement units (IMUs) using Kalman filtering.
 Loosely coupled architectures offer computational simplicity, while tightly coupled methods provide enhanced reliability in GNSS-challenged urban environments \cite{wei2024autonav}. In scenarios where GNSS signals are degraded or unavailable, perception-based localization methods such as Light Detection and Ranging (LIDAR) and visual-inertial simultaneous localization and mapping SLAM have been shown to be promising \cite{9338487}. Radio Detection And Ranging (Radar) systems, particularly frequency-modulated continuous-wave (FMCW) radar, enhance obstacle detection and situation awareness under adverse weather conditions. These onboard systems complement Navigation part by extending perception range and enabling all-weather operability. 
To further improve robustness, multi-sensor fusion architectures integrate data from GNSS, IMU, LiDAR, RADAR, and possibly visual sensors to take advantage of all sensors at the same time.

Surveillance systems such as RADAR and Automatic Dependent Surveillance--Broadcast (ADS-B) remain essential for safety and conformance monitoring. 
Radar plays a critical role in the surveillance infrastructure required for safe and scalable eVTOL operations, particularly in urban environments where visibility may be limited and traffic density is high. Ground-based radar systems have been demonstrated to effectively detect and track non-cooperative aircraft and obstacles, enabling reliable detect-and-avoid capabilities. For instance, NASA’s distributed sensing research has shown that radar, when deployed as part of a networked ground infrastructure, significantly enhances situational awareness by providing persistent coverage, even under poor lighting or weather conditions \cite{nasa2023distributed}. Radar's numerous capabilities and its real-time response make it an indispensable component for low-altitude airspace surveillance in AAM scenarios.

Flight Management Systems (FMS) are foundational components of modern aviation, responsible for automating and optimizing a wide range of flight operations. These systems integrate data from multiple onboard sources, including GPS, INS, air data sensors, and navigation databases, to provide the flight crew with continuous guidance and decision support across all phases of flight. From pre-flight initialization to approach and landing, the FMS facilitates navigation accuracy, fuel efficiency, and workload reduction. FMS functionality is deeply integrated into the avionics architecture of an aircraft. Most systems operate within a closed environment where data flows between the FMS and other subsystems are managed through avionics-specific Local Area Networks (LANs). External communication occurs through structured channels such as the Aircraft Communications Addressing and Reporting System (ACARS), enabling the exchange of flight plans and operational updates between the aircraft and ground-based control centers. These updates are processed through well-defined mechanisms that maintain the consistency and correctness of flight plan data.

Across both domestic and international airspaces, FMS plays an essential role in supporting evolving air traffic management strategies. In the United States, the FAA has introduced the Next Generation Air Transportation System (NextGen), which focuses on enhancing the efficiency and predictability of air traffic through digital coordination and trajectory-based operations. Similarly, the Single European Sky ATM Research (SESAR) initiative in Europe promotes cross-border interoperability and seamless traffic flow using advanced automation and network-centric frameworks. Within these modernized ecosystems, the FMS functions as a key enabler of trajectory management, facilitating dynamic re-routing and real-time performance-based navigation.

As air traffic systems adopt more collaborative and data-driven paradigms, the role of the FMS is also expanding. Ground operation centers increasingly provide initialization data prior to departure, while en route updates can adjust flight trajectories in response to changing conditions such as weather, airspace congestion, or scheduling demands. The integration of these capabilities into the FMS allows for more efficient aircraft turnaround, enhanced situational awareness, and optimized mission execution.

\begin{figure}
    \centering
    \includegraphics[width=1\linewidth]{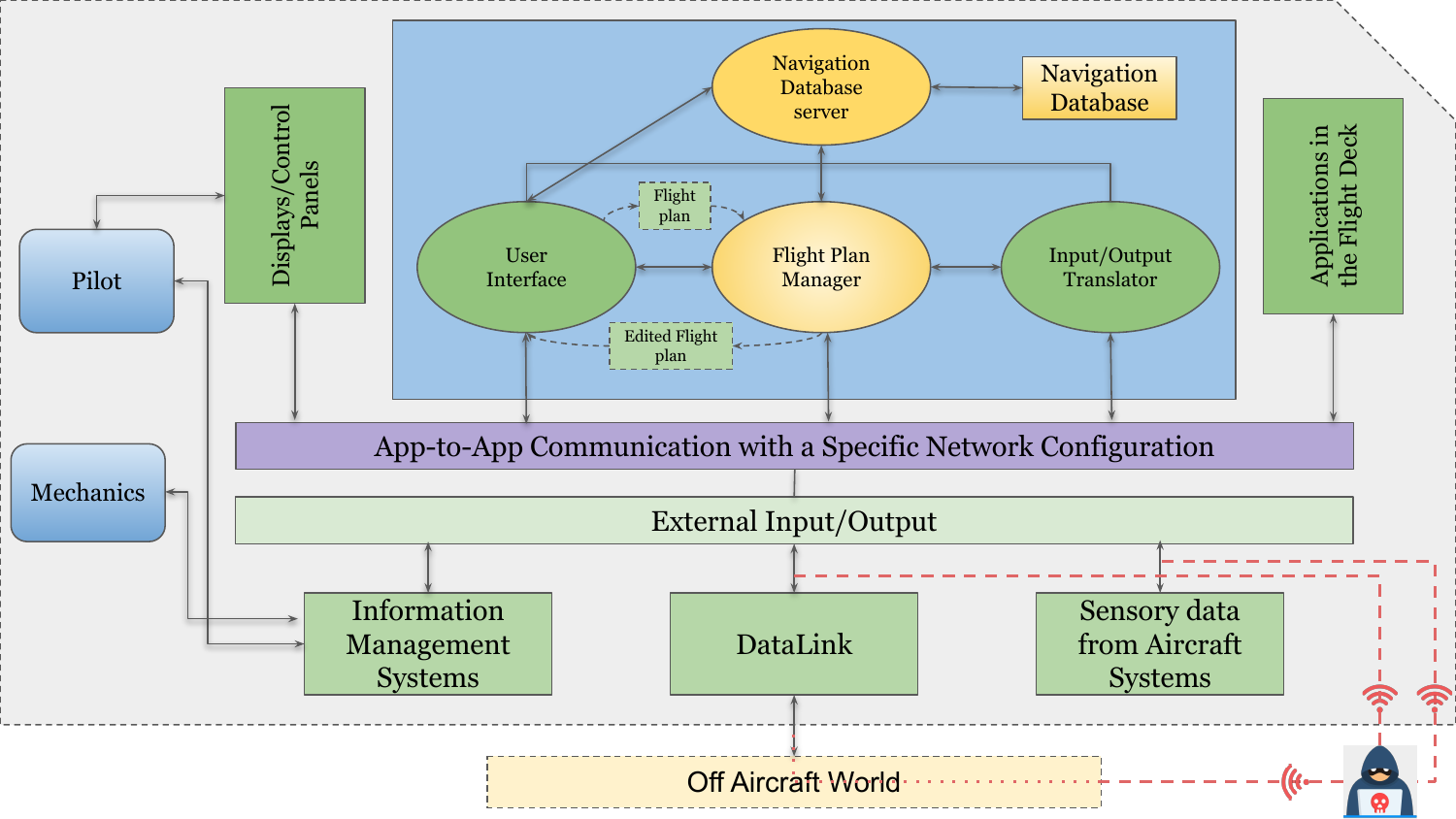}
    \caption{Illustrative architecture of a commercial Flight Management System (FMS),  
      highlighting potential adversarial attack vectors.}
    \label{fig:FMS}
\end{figure}

The general architecture of a modern Flight Management System has been illustrated in figure \ref{fig:FMS}. The architecture presents a high-level schematic showing how the FMS interfaces with both internal aircraft subsystems and external data sources. The FMS is centrally connected to the aircraft's avionic LAN, allowing it to exchange data with navigation sensors, performance monitoring modules, and flight deck displays. Internally, the system receives and processes inputs from the crew and from other aircraft systems to support flight planning, guidance, and optimization functions. As indicated in the diagram, potential adversarial attack vectors include tampering with sensory data inputs or injecting malicious information through datalink communications and other external data sources. These channels, designed to support integration with ground systems, external databases, and application-level interfaces, present opportunities for attackers to compromise flight planning or mislead onboard systems if not properly secured. The diagram also emphasizes the structured nature of these data flows, with clear boundaries between aircraft-internal and external networks.

This system-level integration positions the FMS as a central decision-support component, capable of synthesizing inputs across the aircraft and external airspace infrastructure. As such, the architecture supports both conventional operations and the more dynamic, data-driven environments anticipated in future airspace systems.

\section{Vulnerabilities and Defense Mechanisms} \label{sec: Existing Vulnerabilities and Defense Mechanisms}

\smallskip

AAM and eVTOL systems integrate a diverse set of sensors, communication links, navigation modules, cloud services, and automated flight control components. Each subsystem introduces unique attack surfaces and operational constraints, making the overall ecosystem highly susceptible to multi-vector cyber–physical threats. In the following subsections, we examine the major vulnerabilities across GNSS, communication channels, perception modules in autonomy stacks, avionics functions, data links, and cloud-based UAM service providers. To provide a consolidated view, Table~\ref{tab:threats_defense_threecol} summarizes the key threats identified for each subsystem and outlines the corresponding defense mechanisms discussed throughout this section. Furthermore, Tables~\ref{tab:gnss_threats_defenses}--\ref{tab:fms_supplychain}, as well as Tables~\ref{tab:c2_link_security}--\ref{tab:cloud_aam_risks}, present these vulnerabilities and their proposed solutions in greater detail, together with the associated references.

\begin{table*}[ht!]
\centering
\scriptsize
\caption{Threats and defense mechanisms for key eVTOL / AAM Subsystems in summary.}
\resizebox{\textwidth}{!}{%
\begin{tabular}{|
    >{\centering\arraybackslash\columncolor{RowBlue}}p{0.20\textwidth}|
    >{\raggedright\arraybackslash\columncolor{LightRed}}p{0.38\textwidth}|
    >{\raggedright\arraybackslash\columncolor{LightGreen}}p{0.38\textwidth}|}
\hline
\rowcolor{black!15}
\textbf{Subsystem / Domain} &
\textbf{Existing Threats} &
\textbf{Defense Mechanisms} \\
\hline

\textbf{GNSS / GPS Jamming, Spoofing, NLOS Degradation} &
GNSS denial through jamming; spoofed satellite signals causing incorrect positions; multipath and NLOS degradation; altitude instability; long-duration measurement loss; drift from stale sensor states. &
RTK failover; redundant GNSS references; INS/IMU fusion; LiDAR-, radar-, vision-based navigation; UWB and LEO localization; ML-based spoofing mitigation; robust switching logic; switched-system controllers for GPS-loss states. \\
\hline

\textbf{RF / ATC Voice–Data Impersonation} &
SDR replay of ATC; AI voice cloning; forged digital advisories; masking legitimate transmissions; timing and phraseology anomalies causing unsafe deviations. &
PKI-signed advisories; mutual authentication; RF fingerprinting with transmitter checks; deterministic conflict policy; hold-on-mismatch behavior; EUROCAE secure-voice/data alignment. \\
\hline

\textbf{TCAS / ACAS-X Spoofing and RA Manipulation} &
Ghost aircraft via Mode S / ADS-B spoofing; spurious or conflicting RAs; self-tracking anomalies; oscillatory autopilot responses; trust erosion; vulnerability of 1030/1090 MHz links. &
MLAT cross-check; radar altimeter validation; optical verification; signal-level feature checks (RSSI/TDoA/AoA); trajectory feasibility filters; RA monitoring companions; ATC procedural support and training. \\
\hline

\textbf{ADS-B Spoofing, Identity Manipulation, Saturation} &
Plaintext broadcast enabling spoofing; ghost tracks; identity manipulation; message flooding/DoS; overshadowing/bit-flips; congestion degrading surveillance and DAA. &
MLAT/TDoA verification; onboard sensor fusion; kinematic/TOA filters; congestion-aware track prioritization; confidence scoring via SDSPs; exploration of authenticated ADS-B extensions. \\
\hline

\textbf{Camera and Visual Sensors} &
Adversarial patches hiding aircraft; lighting, weather, and glare degradation; optical deception; multi-sensor adversarial corruption; sensitivity to blur/noise/OOD data. &
LiDAR–camera geometric consistency; distributed sensing; corruption-aware training (AugMix); anomaly detection (Energy/OE); certified fusion architectures; PatchGuard/PatchCleanser; neural-backdoor defenses (Neural Cleanse/STRIP). \\
\hline

\textbf{LiDAR / Radar Sensors} &
False-echo insertion; spoofed returns; range saturation; degraded altitude measurements; cross-sensor adversarial attacks. &
Pulse coding/authentication; multi-band correlation; TOF consistency; redundant-sensor voting; fusion with vision + IMUs for integrity monitoring. \\
\hline

\textbf{Electronic Flight Bag (EFB)} &
Malware on tablets; untrusted Wi-Fi/LTE; tampered performance apps; local database modification; OS vulnerabilities; cross-domain propagation into avionics. &
ML intrusion detection; domain firewalls; signed/versioned content; encrypted sync; certificate-based auth; MDM policy enforcement; rapid rollback; architectural segmentation (AISD/ACD/PIESD). \\
\hline

\textbf{FMS \& Avionics Supply Chain} &
GNSS spoofing corrupting FMS state; ACARS injection; navigation DB tampering; ARINC 664 lateral-movement attacks; supply-chain malware; poisoned cloud routing data; ATM–FMS multi-stage vulnerability propagation. &
Authenticated ACARS; redundant inertial/barometric sources; avionics segmentation; ARINC 664 intrusion detection; STRIDE mitigation; safety-II operational resilience; SBOMs + supply-chain hardening frameworks. \\
\hline

\textbf{Telemetry Data Links} &
Potential spoofing/tampering of fleet telemetry; identity spoofing; rate manipulation; ARINC 429 DoS indirectly suppressing telemetry; cloud-based dependencies. &
Encrypted telemetry; anti-replay freshness windows; segmentation; anomaly monitoring; redundancy + rate limiting; baseline cross-checking of incoming telemetry streams. \\
\hline

\textbf{Navigation Database (Data Attacks)} &
Corruption of vertiport/obstacle/corridor data; poisoned routing; tampering during uplink; ARINC 429 DoS delaying DB transfers; lack of provenance for UAM 3D models. &
Cloud cybersecurity; strict data handling; segmentation; bus-health monitoring; rate-limits; onboard consistency checks; provenance-tracking + integrity auditing (e.g., Merkle-based). \\
\hline

\textbf{C2 Link (Command/Control)} &
Jamming during handovers; MitM; roaming-related latency spikes; spoofed commands; insecure 5G slices; session hijacking; compromised gateways. &
Dual-path diversity (5G + aviation band); spread-spectrum + beamforming; authenticated encryption (TLS/DTLS with AEAD); slice-binding; secure time sources; authenticated control messages; graceful-degradation behaviors. \\
\hline

\textbf{Cloud Architecture / PSUs / SDSP APIs} &
Compromised API credentials; replayed tokens; poisoned SDSP feeds; MitM on cloud channels; stale/incorrect constraints; cascading multi-domain failures. &
OAuth2 + mTLS; signed payloads with freshness; PSU–SDSP cross-validation; immutable audit logs; secure API gateways; provenance tracking; graceful fallback to last-good state. \\
\hline

\end{tabular}%
}
\label{tab:threats_defense_threecol}
\end{table*}

\subsection{Vulnerabilities of onboard avionics and sensors}

The risks and vulnerabilities in this subsection often exist in commercial aviation aircraft, helicopters and general aviation aircraft.

\subsubsection{GNSS/GPS Jamming, Spoofing, and Degradation in Urban Environments}

\begin{table}[t]
    \centering
    \scriptsize
    \caption{GNSS and GPS threat patterns for eVTOL navigation and representative mitigation strategies.}
    \label{tab:gnss_threats_defenses}
    \begin{tabularx}{\linewidth}{%
        >{\raggedright\arraybackslash}p{0.19\linewidth}
        >{\raggedright\arraybackslash}X
        >{\raggedright\arraybackslash}X}
        \toprule
        \rowcolor{black!10}
        \textbf{Threat class} &
        \textbf{Effect on eVTOL operation} &
        \textbf{Representative mitigation} \\
        \midrule

        Jamming of GNSS signals
        & Loss or severe degradation of satellite measurements, position dropouts, unstable altitude estimates, and possible loss of stability in low-altitude UAM corridors \cite{li2023overview,hu2009study,team2014global}. 
        & RTK failover with ground reference stations \cite{boquet2024feasibility}, redundant GNSS sources, and fail-safe switching mechanisms for RTK networks under instability \cite{um2020configuring}. \\[0.2em]

        Spoofing and counterfeit satellites
        & Receiver locks onto fake constellations, computes erroneous positions, and may follow misleading trajectories or be displaced from intended corridors \cite{altaweel2023gps,jafarnia2012gps}. 
        & Multi-sensor fusion with INS, LiDAR, and vision-based navigation to cross-check GNSS solutions \cite{rhudy2012fusion,opromolla2016lidar,petrlik2021lidar,kim2020precise,yol2014vision}; machine-learning-based correction and anomaly detection \cite{zhou2023improvement,brossard2020ai}. \\[0.2em]

        NLOS and multipath in urban canyons
        & Reduced number of usable satellites, biased ranges, degraded vertical position, and long-duration measurement gaps in dense urban environments \cite{hofmann2012global,team2014global}. 
        & Alternative positioning sources such as LEO or cellular signals \cite{8255823,9541006} and UWB hyperbolic navigation \cite{nguyen2016ultra,shule2020uwb}, combined with robust sensor fusion for altitude stabilization. \\[0.2em]

        Long-duration measurement loss
        & Drift of navigation states when controllers reuse stale positions, leading to large deviations or instability. 
        & Using local sensing (IMU, visual, LiDAR) as temporary primary navigation during outages, and designing switched-system controllers that explicitly model GPS-available and GPS-denied modes. \\
        \bottomrule
    \end{tabularx}
\end{table}

For the effective operation of various General Aircraft and eVTOL vehicles utilized in AAM, it is critical that each aircraft can accurately localize its own position (defined as \emph{navigation} in aviation). This level of accuracy is crucial for multiple aircraft to operate simultaneously within the same airspace, since both precise self-localization and real-time awareness of other aircraft positions are necessary to prevent collisions. Unlike conventional aircraft, eVTOLs for air mobility are typically designed to operate in densely populated areas, where they need to fly at lower altitudes and closer to urban environments. Therefore, obtaining and maintaining precise localization signals is crucial for their safe and stable operation. As a result, instability or compromise of positional information (caused by cybersecurity threats such as spoofing, jamming, or various other communication attacks) can lead to severe malfunctions. Most eVTOLs rely heavily on GNSS or GPS signals for their localization. In the GNSS framework, eVTOL aircraft receive signals from multiple satellites and utilize them for geopositioning or navigation through trilateration techniques~\cite{hofmann2012global}. Depending on the country of operation, GNSS systems are known by different names. For instance, GPS is operated by the United States, Beidou by China, GLONASS by Russia, and Galileo by the European Union. For simplicity, this section will collectively refer to these systems as GNSS or GPS. In positioning via GNSS, the accuracy of the calculated location increases with the number of satellite signal sources utilized. Therefore, to ensure a stable and precise GNSS-based positioning system, it is crucial to guarantee the secure reception of these signals.

\paragraph{Existing Threats:} Accurate reception of GNSS signals is crucial for the localization of eVTOLs. The performance of the system is clearly vulnerable the unavailability of a (GPS) signal for synchronization, whether due to urban canyons or intentional jamming. In the event of such measurement losses, eVTOL aircraft may suffer from poor performance and potentially lose stability or display unacceptable performance otherwise. There exist multiple ways to attack the signals received by a vehicle, with GPS jamming being the most representative. GPS jamming~\cite{li2023overview,hu2009study} refers to an attack technique similar to signal jamming, in which specific radio signals are transmitted to overwhelm GPS satellite signals. Through such interference, the GNSS signals received by the vehicle can be blocked or destabilized, significantly degrading positional accuracy and ultimately compromising localization and navigation capabilities. Beyond jamming, GPS spoofing~\cite{altaweel2023gps,jafarnia2012gps} represents another method of undermining GNSS-based localization. Signal spoofing deceives the GNSS receiver by generating counterfeit satellite signals, thereby injecting false positioning data. As a result, the vehicle calculates its position based on inaccurate information, leading to erroneous localization that can cause severe system failures or even enable hijacking scenarios.
In addition to these external attacks, environmental factors can also degrade the accuracy of GNSS systems. GNSS precision deteriorates when a vehicle operates in a Non-Line-of-Sight (NLOS) environment, where the direct path between the satellite and the vehicle is obstructed. For instance, operations in dense urban areas surrounded by tall buildings, or in valleys or mountainous regions, significantly reduce the likelihood of receiving Line-of-Sight (LOS) GPS signals, making stable signal acquisition difficult and thus leading to degraded GNSS performance. In such cases, signal blockage caused by the surrounding environment prevents the receiver from obtaining accurate data, ultimately lowering overall GPS positioning accuracy. Among the positional coordinates, altitude values are known to be particularly unstable compared with latitude and longitude~\cite{team2014global}. While the latter can be corrected or stabilized using supplementary sensor data, inaccuracy in altitude measurements poses a critical risk. This altitude uncertainty can be especially hazardous for eVTOL operations within UAM environments, where precise vertical positioning is essential for safe navigation and separation management.
Therefore, although GNSS/GPS remains a fundamental component in eVTOL localization, reliance on a single source of positioning information renders the system highly susceptible to the aforementioned cybersecurity and environmental vulnerabilities. To enhance system robustness, it is imperative to integrate complementary navigation and localization methods, thereby achieving greater resilience and reliability.

\paragraph{Defense Mechanisms:} To address the vulnerabilities of GNSS/GPS-based positioning and navigation systems, a naïve approach is to allow the controller to use the last available measurement in the event of a measurement loss, and bound the maximum allowable duration of the loss to guarantee stability. However, the duration of measurement loss may easily exceed the maximum allowable duration to guarantee stability due to the fact aircraft dynamics are extremely fast and may therefore not be able to maintain stability under the longer measurement loss durations in GPS signal losses. Moreover, even for measurement loss durations smaller than this threshold, large position and velocity drifts can occur due to the controller repeatedly using the incorrect (last available) measurement. Therefore, multiple studies have been conducted to improve the robustness of the GNSS/GPS system while addressing these issues. First, research has been carried out to enhance fail-safe mechanisms in cases of GPS signal failure. To compensate for the instability of GPS signal reception, the Real-Time Kinematic (RTK,~\cite{boquet2024feasibility}) system has been developed. This method utilizes a stable GNSS signal received at a stationary base station to correct unreliable GPS values and improve accuracy. Based on this concept, \cite{um2020configuring} proposed fail-safe switching mechanisms that ensure redundancy in RTK-GPS during network instabilities caused by various threats. Next, some studies focus on integrating multiple sensors to mitigate navigation errors arising from GPS vulnerabilities. For instance, alternative or supplementary sensors can be utilized to support or replace conventional GPS-based navigation under such threat conditions. These include the use of Inertial Navigation Systems (INS) employing IMU sensors~\cite{rhudy2012fusion}, localization and navigation using LiDAR sensors~\cite{opromolla2016lidar,petrlik2021lidar}, improving altitude and positional accuracy through vision sensors~\cite{chen2025obtpn,kim2020precise,yol2014vision}, backup navigation using Low Earth Orbit (LEO) satellites or cellular signals~\cite{8255823, 9541006}, and hyperbolic navigation methods using Ultra-Wideband (UWB) signals~\cite{nguyen2016ultra, shule2020uwb}. Finally, various machine learning techniques have also been applied to enhance positioning accuracy under potential GPS threats. For example, Recurrent Neural Network~\cite{zhou2023improvement} or Convolutional Neural Network~\cite{brossard2020ai} can be employed to improve the accuracy of GPS signals by compensating for inaccuracies that arise during operation.

\paragraph{Gaps and Open Problems:} Despite substantial studies and notable progress in developing robust GNSS-based localization systems for eVTOLs, there is still no widely accepted or standardized integrated framework to address this issue. Instead, many researchers and operators rely on their own proprietary or ad-hoc methods to ensure the navigation accuracy of GNSS systems. Furthermore, variations in international and organizational regulations have hindered the establishment of lightweight civil-signal authentication mechanisms for GNSS suited for eVTOL applications. In addition, there remains a lack of certifiable machine learning–based jamming and spoofing detectors implemented in practical settings. 

For robustness to improve the reliability of the system, there needs to be a process by which when the GPS measurement is lost at an eVTOL aircraft, local sensing (based on IMU and visual sensors, or signal triangulation from LEO with respect to a pre-fed map of landmarks) is temporarily used for primary control in place of GPS signal to provide navigation. One complication here is that each eVTOL may operate with GPS-based or local-sensing based control at any time instant, depending on the availability of GPS measurements at that aircraft. More research is needed on how to analyze the resulting system and control architecture as a nonlinear switched system and possibly design distributed controllers to guarantee safety of all the aircraft in the same airspace.

\subsubsection{RF/ATC Link Impersonation and Unauthorized Frequency Use}

\begin{table}[t]
\centering
\scriptsize
\caption{RF and ATC link impersonation in AAM corridors: attack modes and protection layers.}
\label{tab:rf_atc_impersonation}
\begin{tabularx}{0.97\linewidth}{@{}X@{}}
\toprule

\textbf{Voice replay on aeronautical channels} \\
\midrule
\textbf{Capability:} Records and replays authentic ATC clearances with correct phraseology \cite{faaUAM2023}. \\
\textbf{Symptom:} Duplicate or conflicting clearances, unexpected altitude/corridor changes. \\
\textbf{Defense:} Query–hold policies; cross-check against signed digital advisories. \\[0.5em]
\addlinespace[0.4em]
\toprule

\textbf{AI voice cloning of controllers} \\
\midrule
\textbf{Capability:} Real-time imitation of ATC voices using neural voice-cloning models \cite{voice-cloning}. \\
\textbf{Symptom:} Highly realistic but unauthorized instructions consistent with prior interactions. \\
\textbf{Defense:} RF fingerprinting, TOA validation of transmitters \cite{rf-fp-review,rf-fp-vs-ia}; pilot training for anomaly detection. \\[0.5em]
\addlinespace[0.4em]
\toprule

\textbf{Carrier interference and masking} \\
\midrule
\textbf{Capability:} Transmits noise or blocking signals on the ATC frequency \cite{rfi-aviation}. \\
\textbf{Symptom:} Broken / clipped transmissions, missing clearances in dense UAM corridors. \\
\textbf{Defense:} Channel monitoring, automatic fallback channels, deterministic loiter-on-silence procedures \cite{AAM-implementation-plan,faaUAM2023}. \\[0.5em]
\addlinespace[0.4em]
\toprule

\textbf{Coordinated voice + data forgery} \\
\midrule
\textbf{Capability:} Spoofed digital advisories aligned with forged voice commands. \\
\textbf{Symptom:} Very high plausibility due to synchronized voice–data deception. \\
\textbf{Defense:} Mutual TLS, PKI, anti-replay for digital advisories \cite{eurocae-info-sec-vtol}; signed data overrides voice. \\

\bottomrule
\end{tabularx}
\end{table}

In the United States and other countries,
government bodies such as the Federal Communications Commission (FCC) and National Telecommunications and Information Administration (NTIA) regulate the radio frequency (RF) spectrum and reserve certain frequency bands for aeronautical mobile communications,
such as those between eVTOL crews and remote supervisors or air traffic control.
Anyone with suitable radio equipment has the capability to broadcast on a given RF frequency band,
meaning that an adversary can transmit on the same aeronautical voice or ground radio channels used by eVTOL crews or remote supervisors
and imitate a controller or vertiport dispatcher.
In the absence of defense mechanisms, an adversary may fabricate messages, replay prior clearances, or splice phrases timed to blend with legitimate traffic.
In dense, low-altitude corridors with high tempo operations, such forgeries can be wrongly accepted, producing off-nominal trajectory changes (wrong corridor/altitude or incorrect vertiport turn-in).
This risk is recognized for UAM voice/data exchanges and human–automation coordination in FAA UAM ConOps v2.0~\cite{faaUAM2023}.

\paragraph{Existing Threats:}
Since broadcast transmissions from entities are unencrypted by default,
an adversary may record transmissions and use them to construct an attack.
A simple example would be a voice replay attack where the adversary transmits a previously recorded clearance message from ATC
with correct phraseology and timing to make an eVTOL crew believe they have clearance when they should not.
A more sophisticated version of this would be to use AI voice cloning technology~\cite{voice-cloning} to mimick the voice of a controller in real-time.
An adversary may also listen to transmissions to determine a time when they want to transmit an unauthorized carrier on some channel
as a form of RF interference~\cite{rfi-aviation} to mask or preempt legitimate transmissions,
preventing a given entity from receiving important information.
Since operators often trust corroboration between voice and data channels,
an even more sophisticated attack would be to simultaneously forge a voice transmission and digital advisory to increase plausibility.
These attacks can have observable symptoms and operational impacts such as
timing irregularities, atypical phraseology, or RF fingerprints inconsistent with authorized ground transmitters.
They can result in loss of separation or missed approach sequencing in corridors or vertiports as noted in AAM concepts of operation~\cite{AAM-implementation-plan}.

\paragraph{Defense Mechanisms:}
An obvious defense mechanism for digital communications is the use of cryptography.
Mutual authentication and freshness for digital advisories could be achieved through the use of Mutual Transport Layer Security (mTLS) and public key infrastructure (PKI) with cryptographic nonces and timestamps to prevent replay attacks,
aligning with EUROCAE VTOL information security guidance~\cite{eurocae-info-sec-vtol}.
For analog communications,
RF transmitter fingerprinting~\cite{rf-fp-review, rf-fp-vs-ia}
through monitoring carrier, time of arrival (TOA), and in-phase and quadrature (IQ) artifacts and comparing them to known transmitter locations for ground stations
can provide a means of identifying impersonated communications,
subsequently triggering query/hold policies.
When comparing digital and analog communications,
a deterministic conflict policy can be applied such that,
if voice and signed data disagree,
the eVTOL can default to hold/loiter until the conflict is resolved.
This approach would be consistent with safety-first fallbacks in the UAM ConOps~\cite{faaUAM2023}.

\paragraph{Gaps and Open Problems:}
Interoperable secure voice and data procedures tailored for AAM are still maturing.
There are limited urban-multipath radio voice datasets for spoofing and anomaly detection,
and there is limited certification guidance for time-critical RF communication~\cite{faaUAM2023, eurocae-info-sec-vtol}.

\keptcontentend


\subsubsection{TCAS and ACAS\,-X Vulnerabilities}

\begin{table}[t]
    \centering
    \scriptsize
    \caption{TCAS and ACAS-X security-relevant layers, failures, and mitigation concepts.}
    \label{tab:tcas_acas_layers}

    \begin{tabularx}{0.85\linewidth}{|
        p{0.16\linewidth}|
        X|
        X|
        X|}
    \hline
    \multicolumn{4}{|c|}{\cellcolor{black!10}\textbf{ACAS-X Security and Vulnerability Overview}} \\
    \hline

    \textbf{Layer} &
    \textbf{Role in TCAS / ACAS-X} &
    \textbf{Security vulnerability} &
    \textbf{Mitigation direction} \\
    \hline

    Surveillance inputs &
    Mode S and ADS-B replies on 1030/1090 MHz provide range, altitude, and closure rate for intruder aircraft \cite{Eurocontrol2017ACASGuide,FAA2011TCASIntro}. &
    Unauthenticated messages enable ghost tracks, replay attacks, and self-tracking anomalies, which can trigger spurious RAs \cite{habler2023assessing,cisa2025tcas}. &
    Multi-source validation using MLAT and radar, plus signal-level checks (RSSI, TDoA, AoA) to flag inconsistent or physically impossible tracks \cite{strohmeier2020securing,kuba2024navigating}. \\[0.6em]
    \hline

    Decision policy (ACAS-X tables) &
    Markov decision process optimized tables select RAs that balance safety and operational costs \cite{kochenderfer2012next,guendel2025collision}. &
    Tables assume honest surveillance; adversarial trajectories can drive the system into unnecessary or conflicting advisories. &
    Companion monitoring of RA histories, trajectory feasibility checks, and future formal verification under adversarial input models \cite{smith2022understanding,kuba2024navigating}. \\[0.6em]
    \hline

    Flight deck integration &
    Advisories appear on cockpit displays and may be coupled to the autopilot for automatic execution \cite{harman1989tcas,smith2022understanding}. &
    Coupling to autoflight can produce oscillatory behavior or large altitude deviations when corrupted inputs persist; repeated false alerts reduce crew trust \cite{smith2020view,FAA_AC90_120}. &
    Training and procedures for recognizing abnormal alert patterns, guidance for reverting to TA-only modes, and joint ATC–airline reporting workflows for anomaly investigation \cite{FAA_AC90_120,habler2023assessing}. \\[0.6em]
    \hline

    Operational ecosystem &
    TCAS and ACAS-X form the last airborne safety net on top of pilot vigilance and ATC separation \cite{smith2022understanding}. &
    Lack of message authentication and incomplete standards for secure surveillance keep CAS exposed to targeted spoofing campaigns. &
    Development of lightweight authentication or integrity protection for surveillance channels, co-designed with CAS logic and certification constraints. \\
    \hline

    \end{tabularx}
\end{table}


Collision Avoidance Systems (CAS) represent one of the most critical airborne safety mechanisms, designed to reduce the risk of mid-air collisions by continuously monitoring the surrounding airspace and issuing maneuvering instructions when the risk of collision becomes imminent \cite{smith2022understanding}. These systems operate independently of ground-based infrastructure and use transponder replies from nearby aircraft to determine range, altitude, and closure rate \cite{Eurocontrol2017ACASGuide}. The CAS logic continuously evaluates time to closest point of approach and generates alerts when separation thresholds are predicted to be violated \cite{smith2022understanding}. Such airborne self-separation capabilities serve as a final protective layer in the air traffic management hierarchy, supplementing both pilot vigilance and ATC separation procedures \cite{smith2022understanding, harman1989tcas}.

The Traffic Collision Avoidance System (TCAS) \cite{harman1989tcas}, standardized as ACAS~II by the International Civil Aviation Organization (ICAO), constitutes the current operational realization of CAS in commercial aviation. TCAS operates independently of ground-based air traffic control by using transponders to interrogate the airspace around an
aircraft \cite{FAA2011TCASIntro}. TCAS performs active surveillance by transmitting Mode~S interrogations on 1030~MHz and receiving replies on 1090~MHz from aircraft equipped with transponders \cite{smith2022understanding}. Based on closure rate, altitude difference, and vertical speed, TCAS issues \textit{Traffic Advisories (TAs)} and \textit{Resolution Advisories (RAs)} that instruct the pilot to climb or descend to maintain safe separation \cite{smith2022understanding, harman1989tcas, habler2023assessing}. These advisories are presented on cockpit displays and, in advanced implementations, can be linked with the flight director or autopilot to enable automatic execution \cite{habler2023assessing}. The system has been highly successful in reducing mid-air collisions, but its reliance on unauthenticated transponder data remains a critical security concern \cite{habler2023assessing, smith2022understanding}.

\begin{figure}
    \centering
    \includegraphics[width=1\linewidth]{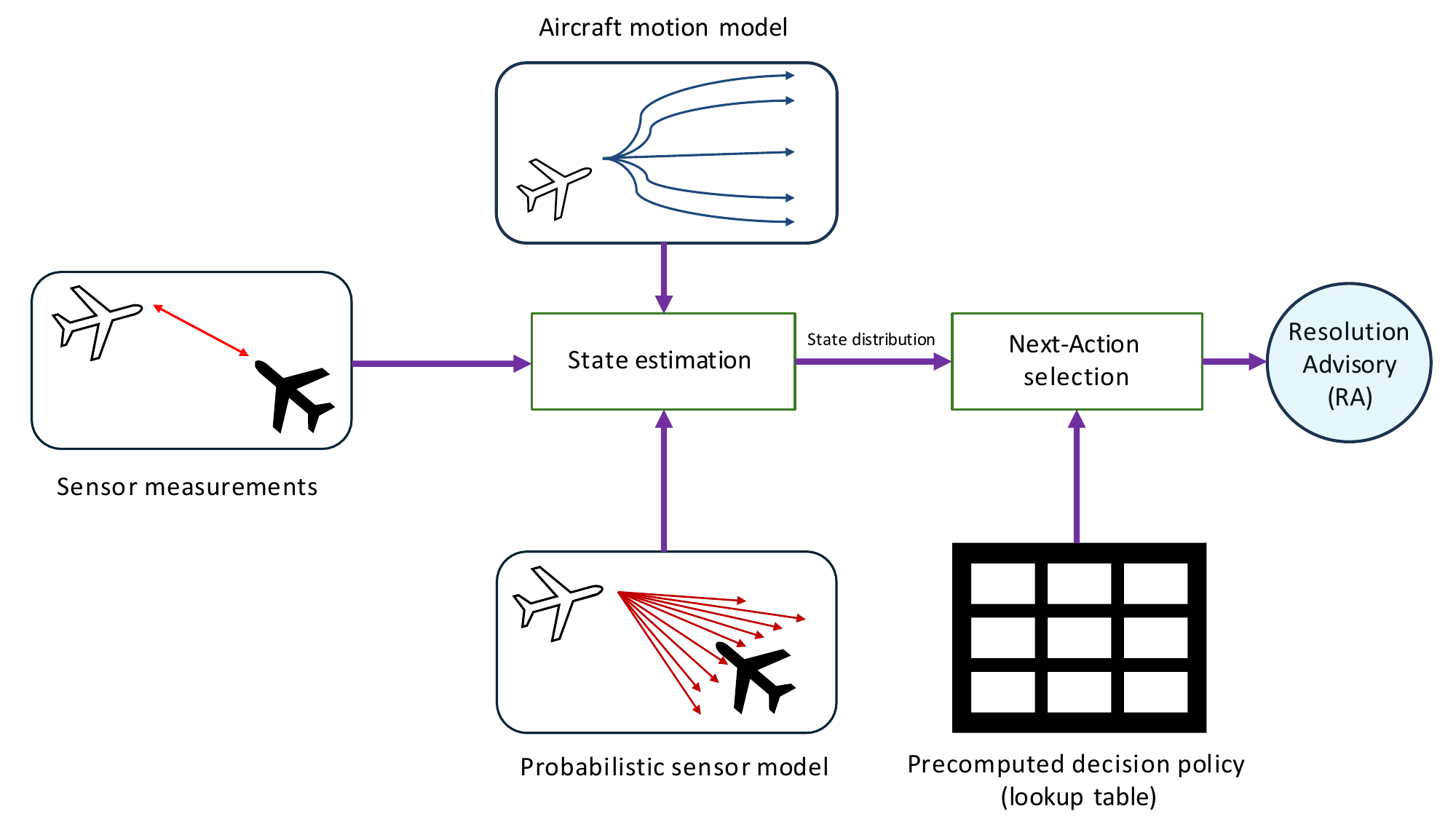}
    \caption{ACAS-X continuously updates its aircraft state estimates as new sensor measurements become available. The system represents uncertainty using a probability distribution, which determines the appropriate region of the lookup table from which to select the corresponding resolution advisory.}
    \label{fig:ACASX}
\end{figure}

The Airborne Collision Avoidance System~X (ACAS-X) represents the next generation of collision avoidance technology, developed under a joint FAA–MIT~Lincoln Laboratory program to address TCAS limitations in logic efficiency and operational acceptability \cite{kochenderfer2012next}. ACAS-X retains the existing surveillance infrastructure (Mode~S, ADS-B) but replaces the rule-based logic tables of TCAS with a decision policy obtained through large-scale Markov decision process optimization \cite{kochenderfer2012next}. Its alerting tables are derived offline from probabilistic encounter models that balance safety metrics and operational costs, significantly reducing unnecessary advisories and alert reversals \cite{Eurocontrol2017ACASGuide}. Figure \ref{fig:ACASX} shows the architecture of the proposed ACAS-X. Multiple variants have been proposed, such as ACAS~Xa for manned aircraft, ACAS~Xu for unmanned systems, and ACAS~Xr for rotorcraft \cite{guendel2025collision}, extending its applicability to future AAM and eVTOL operations.

\paragraph{Existing Threats:}
While ACAS-X enhances operational robustness compared to TCAS, it inherits several fundamental vulnerabilities from the underlying surveillance infrastructure. Both systems depend on transponder-based inputs transmitted over unauthenticated 1030/1090~MHz channels (Mode~S, Secondary Surveillance Radar (SSR), ADS-B), which are susceptible to spoofing, replay, and signal injection. Prior studies such as \cite{cisa2025tcas} have demonstrated that adversaries can fabricate synthetic aircraft tracks through manipulated ADS-B or Mode~S transmissions, leading to erroneous threat evaluations and, in some cases, the generation of unnecessary or conflicting RAs. Experiments conducted using standardized ACAS-X code and real traffic datasets showed that such false inputs can trigger spurious advisories in approximately 44~percent of the simulated encounters, increasing to nearly 79~percent within the altitude range of 2,350–13,300~ft~AGL. The induced deviations averaged roughly 590~ft, indicating a significant operational impact even when the attack is limited to radio-layer manipulation \cite{smith2022understanding}.

Beyond direct signal spoofing, the interaction between TCAS or ACAS-X and other avionics subsystems introduces additional pathways for exploitation. The exchange of data among the transponder, display units, and autoflight systems can propagate corrupted inputs or stale advisories, amplifying the effects of an external disturbance. The coupling between TCAS and autopilot functions, particularly when RAs are executed automatically, can result in hazardous oscillations or mode confusion when multiple systems respond simultaneously to falsified data \cite{kuba2024navigating}. Moreover, sustained false or nuisance alerts have been observed to erode pilot trust, prompting flight crews to revert to TA-ONLY or STBY modes, thereby disabling the last line of airborne collision protection \cite{smith2020view}. Notably, FAA operational guidance acknowledges that ACAS~II and ACAS~Xa may generate false or self-tracking RAs due to surveillance or tracking anomalies, including instances where an aircraft may momentarily track itself as a threat. These spurious advisories can lead to large altitude deviations if not properly identified, and flight crew are explicitly instructed to report such anomalies for post-event safety investigation \cite{FAA_AC90_120}. As highlighted in previous analyses, the absence of message authentication and the reliance on assumed integrity in Mode~S replies remain the principal attack vectors for CAS technologies across both legacy and next-generation implementations \cite{habler2023assessing}.

\paragraph{Defense Mechanisms:}
Several mitigation strategies have been proposed to reduce the exposure of TCAS/ACAS-X and related collision avoidance systems to spoofing and signal manipulation. One class of defenses focuses on enhancing the integrity of surveillance data through multi-source validation. By cross-referencing ADS-B or Mode~S inputs with independent sensors such as multilateration (MLAT) \cite{strohmeier2020securing}, radar altimeters, or optical detection systems, it becomes possible to identify injected or inconsistent tracks before they influence the decision logic. Research efforts have shown that incorporating signal-level characteristics, including received power, time difference of arrival, or angle-of-arrival estimates, can substantially improve the ability to distinguish genuine aircraft replies from fabricated ones \cite{kuba2024navigating}. Additionally, These parameters can be used to compute reliability estimates for intruder tracks or to reject trajectories that violate physical feasibility, thereby preventing false alerts caused by manipulated or corrupted surveillance inputs.
From an architectural standpoint, layered monitoring systems have been proposed to operate in parallel with ACAS-X logic. These companion monitors would evaluate real-time trajectory consistency and alert reliability, flagging sequences of repeated false RAs that may indicate a spoofing or replay campaign \cite{kuba2024navigating}. Furthermore, procedural defenses such as standardized crew responses, ATC coordination protocols, and training for recognition of abnormal alert patterns have been recommended to complement technical safeguards \cite{habler2023assessing}.

\paragraph{Gaps and Open Problems:}
Although ACAS-X provides measurable improvements in alerting logic, several open research and certification challenges persist. The foremost gap concerns the absence of authenticated or integrity-protected surveillance inputs. No globally standardized method currently exists to validate Mode~S or ADS-B messages within the latency and interoperability constraints of commercial avionics. As a result, ACAS-X decision policies continue to rely on unverified sensor data, leaving the system vulnerable to targeted signal manipulation. Another critical issue lies in the formal verification of the ACAS-X decision policy under adversarial or degraded conditions. Existing validation processes focus on nominal encounter models and do not account for maliciously crafted input trajectories that could exploit the decision table boundaries or cost-function discontinuities. Furthermore, the growing degree of integration between ACAS-X advisories and automated flight control systems raises concerns regarding closed-loop stability and pilot situational awareness in the presence of false or oscillatory RAs. Addressing these gaps will require the development of certified, lightweight authentication mechanisms for surveillance data, extended verification frameworks incorporating adversarial inputs, and human-factors studies focused on maintaining trust and compliance under abnormal alerting conditions.


\subsubsection{ADS\,-B Attacks and Spectrum Saturation}

\begin{table}[t]
\centering
\scriptsize
\caption{ADS-B misuse patterns and defense principles for high-density AAM corridors.}
\label{tab:adsb_security}

\begin{tabularx}{0.98\linewidth}{
    >{\raggedright\arraybackslash}p{0.20\linewidth}
    !{\vrule width 0.8pt}
    >{\raggedright\arraybackslash}p{0.33\linewidth}
    >{\raggedright\arraybackslash}X
}
\toprule
\rowcolor{black!10}
\textbf{Attack Type} &
\multicolumn{1}{c}{\textbf{Impact on Surveillance / DAA}} &
\multicolumn{1}{c}{\textbf{Defense Strategy (Refs.)}} \\
\midrule
\rowcolor{black!5}
\textbf{Ghost track injection} &
Fabricated aircraft clutter displays and mask real conflicts
\cite{pearce2021signal,slimane2022ads,khan2024survey}. &
Cross-validate ADS-B using MLAT (TDoA), LiDAR–inertial localization, or optical sensing to confirm track legitimacy
\cite{rhudy2012fusion,opromolla2016lidar,petrlik2021lidar}. \\
\midrule

\textbf{Identity manipulation} &
Spoofed Mode S/ADS-B identifiers disrupt conformance monitoring and separation
\cite{shang2019multidevice,khan2024survey}. &
Identity-integrity scoring, track-continuity validation, and cooperative confidence sharing between PSUs/SDSPs
\cite{chen2025obtpn}. \\
\midrule
\rowcolor{black!5}

\textbf{Flooding / 1090 MHz congestion} &
Message flooding saturates channel capacity, degrading surveillance reliability
\cite{popper2011investigation,khan2024survey}. &
Congestion-aware filtering prioritizing nearby/conflict-relevant targets; rate limiting and deprioritizing low-confidence tracks
\cite{kim2020precise}. \\
\midrule

\textbf{Plaintext broadcast} &
Unauthenticated broadcast enables eavesdropping and low-cost spoofing with SDRs
\cite{pearce2021signal,slimane2022ads}. &
Research toward authenticated ADS-B extensions balancing scalability, confidentiality, and backward compatibility. \\
\bottomrule

\end{tabularx}
\end{table}

The Automatic Dependent Surveillance–Broadcast (ADS-B) system, mandated by civil aviation authorities worldwide, is a cornerstone of modern air traffic management. It enables each aircraft to periodically broadcast its identity, position, velocity, and operational status, thereby supporting surveillance, conformance monitoring, and detect-and-avoid (DAA) functions. The simplicity of message decoding and the availability of low-cost software-defined radios (SDRs) have further accelerated ADS-B adoption in both civil and unmanned aviation domains.

However, ADS-B’s open and unauthenticated broadcast nature makes it inherently vulnerable to cyberattacks. Because messages are transmitted in plaintext without encryption or authentication, adversaries can intercept, modify, or fabricate signals to inject false tracks or manipulate aircraft identifiers. Past research works have demonstrated that injection, identity manipulation, and flooding attacks can be executed using inexpensive SDR equipment \cite{pearce2021signal,slimane2022ads,shang2019multidevice}. In dense or low-altitude UAM environments, where surface or corridor surveillance may fuse ADS-B tracks for conformance and DAA, these injected or duplicate tracks can obscure genuine conflicts and trigger nuisance advisories, undermining situational awareness and airspace safety \cite{khan2024survey}. Additional vulnerabilities include denial-of-service (DoS) through message flooding or spectrum congestion, message modification via overshadowing or bit-flipping, and jamming of the 1090~MHz frequency band used by ADS-B \cite{popper2011investigation}. As urban airspace becomes increasingly populated with eVTOL operations, spectrum saturation may happen. Therefore, the ADS-B system, while critical for UAM operation, exposes a significant cyber attack surface for both intentional and unintentional interference.

\paragraph{Defense Mechanisms:}
Effective defenses against ADS-B exploitation combine data cross-validation, signal-level filtering, and adaptive prioritization.  
First, cross-validation techniques correlate ADS-B reports with independent surveillance or onboard sensing modalities, such as multilateration (MLAT) using time-difference-of-arrival (TDoA), LiDAR–inertial localization, or optical perception systems, to verify the plausibility of received tracks \cite{rhudy2012fusion,opromolla2016lidar}. Second, kinematic and time-of-arrival (TOA) filtering rejects tracks that exhibit implausible dynamics, inconsistent timestamp patterns, or physically infeasible motion signatures \cite{petrlik2021lidar}.  
Finally, priority management under congestion ensures that surveillance and traffic management systems allocate bandwidth and processing resources to nearby or conflict-relevant tracks, while rate-limiting or deprioritizing distant or low-confidence data. These confidence scores can be shared between service providers, such as Provider of UTM Services (PSU) or Supplemental Data Service Providers (SDSP), to maintain common situational awareness under saturation conditions \cite{chen2025obtpn}.

\paragraph{Gaps and Open Problems:}
Despite ongoing research, there remains no globally adopted framework for authenticated yet privacy-preserving ADS-B communication suited to high-density UAM operations. Current cryptographic proposals face scalability and backward-compatibility challenges, while physical-layer cross-validation depends heavily on network geometry and sensor availability. Moreover, public, vendor-neutral trials quantifying ADS-B resilience under urban-density conditions are still limited. Future work should focus on certifiable, privacy-aware surveillance protocols and empirical testing in representative eVTOL and UAM corridors to evaluate the effectiveness of saturation and spoofing defenses \cite{kim2020precise}.

\subsubsection{Electronic Flight Bag (EFB)}


Electronic Flight Bags (EFBs) have evolved as essential digital tools that replace traditional paper-based flight materials, providing flight crews with electronic access to navigation charts, operational manuals, and aircraft checklists \cite{ukwandu2022cyber}. Initially implemented as laptop-based systems performing operational and flight management functions, EFBs now accommodate a wide range of software applications that automate tasks traditionally carried out manually, such as takeoff performance, weight and balance, and landing calculations \cite{melnichuk2019development}. Their adoption by commercial airlines, supported by early digital operations suites offered by major avionics manufacturers, has demonstrated measurable gains in operational efficiency and flight safety \cite{bhardwaj2019safety, samosir2021effect}. Compared with standard laptop software, EFBs offer the advantage of not requiring separate storage procedures below 10,000 feet, which simplifies in-flight usage \cite{mecham2002new}.

Modern EFBs are designed to enhance situational awareness, safety, and efficiency by integrating seamlessly with the FMS and other avionics to display an aircraft’s real-time position, weather overlays, and relevant operational data \cite{ukwandu2022cyber,marinos2020aviation}. \cite{foreflight2025mobile} shows example screens from a commercial EFB application, illustrating typical functionality for flight planning, procedure display, weather visualization, and digital checklists. Advanced configurations, including Class~3 EFB systems, enable integration with (ADS-B) and polarimetric radar technologies, further supporting route optimization and improved pilot decision-making \cite{lupidi2015contributing,zelazo2012electronic}. These digital systems also facilitate faster information exchange between the cockpit and ground support, streamlining flight operations and business processes.






Research indicates that EFBs generally improve flight crew performance by providing quick access to preflight information without significantly increasing workload under normal conditions \cite{chandra2003human}. However, studies such as \cite{suppiah2019impact,lopes2022supporting} report that during unexpected or high-stress scenarios, EFB interactions can temporarily elevate pilot workload. Despite these operational trade-offs, EFBs remain a cornerstone of modern cockpit digitalization, offering substantial advantages in safety, efficiency, and accessibility over conventional paper-based methods. 

\paragraph{Existing Threats:}

\cite{wolf2014information} emphasizes that EFB enhances efficiency and reduces onboard weight compared to paper-based systems. However, the connectivity that enables these advantages also expands the attack surface: a compromised or malware-infected EFB could serve as an entry point for denial-of-service or similar attacks against interconnected avionics systems. \cite{habler2023assessing} demonstrates that this interconnectivity creates a pathway for adversaries to pivot between passenger, administrative, and control networks if domain separation fails. As commercial off-the-shelf tablets are increasingly used as Class~2 EFBs, they inherit vulnerabilities from consumer ecosystems, including malicious applications, overheating-induced shutdowns, software crashes, and unverified data updates \cite{bhardwaj2019safety}. Furthermore, EFB reliance on wireless connectivity for database synchronization and NOTAM updates introduces exposure to spoofed or manipulated data transfers, echoing the broader avionics threat surface observed in FMS and ACARS links \cite{liu2024optimal,habler2023assessing}. The portable nature of the EFB and its ability to connect to public networks put the device at high risk, since exposure to open Wi-Fi can permit remote exploitation through unpatched software vulnerabilities \cite{HowSecureAreIFECSystems2017}. 

\begin{figure}
    \centering
    \includegraphics[width=1\linewidth]{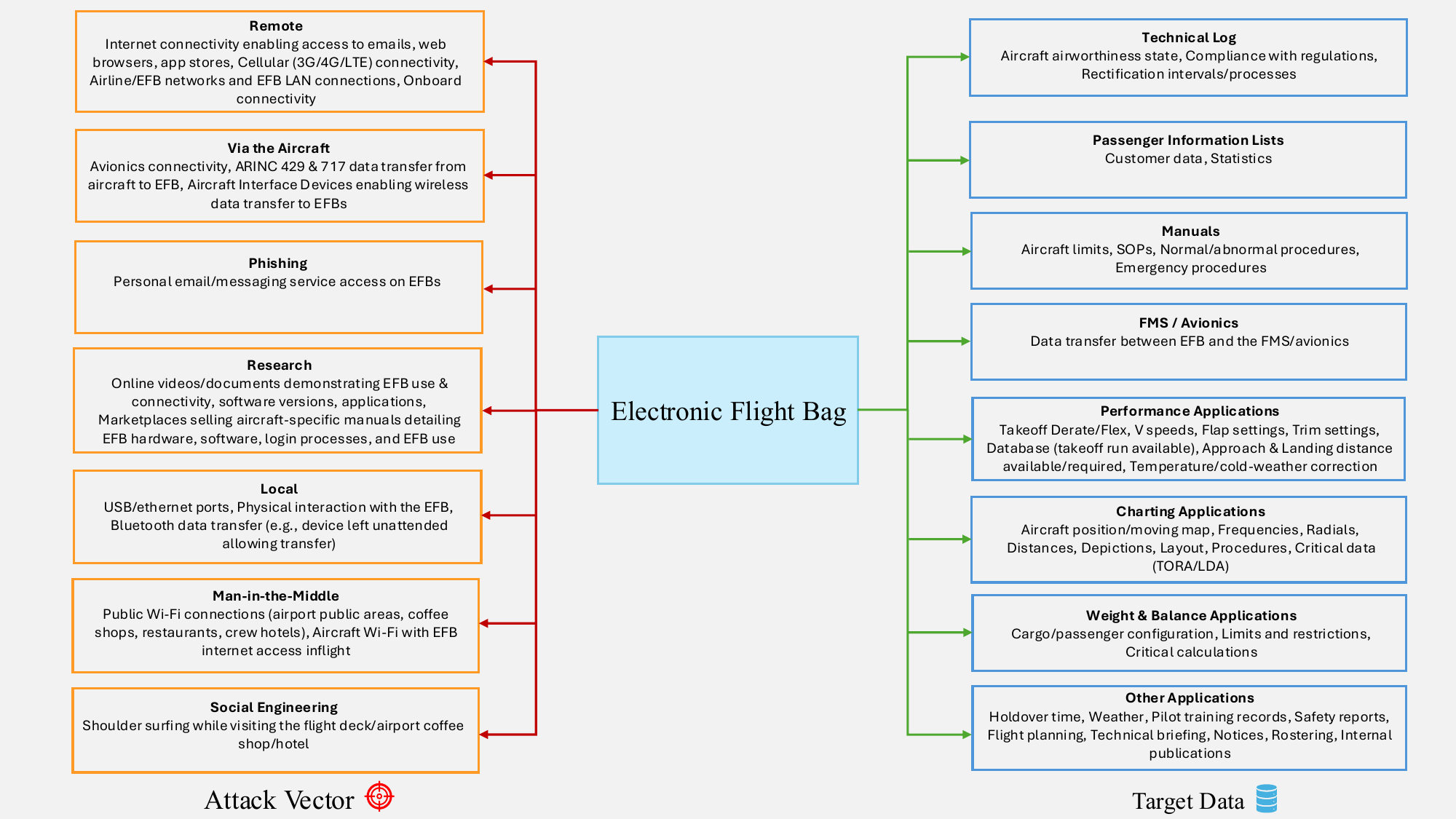}
    \caption{Attack vectors and target data categories for EFBs, highlighting how operational, technical, and passenger-related data may be compromised through remote, local, or human-based threats.}
    \label{fig:EFB}
\end{figure}

Recent investigations highlight additional operational and cyber risks unique to EFB deployments. According to \cite{pentestpartners2023efb1}, inconsistencies in EFB classification and certification standards among operators lead to uneven security postures, where some devices are treated as portable consumer tablets rather than regulated avionics components. \cite{pentestpartners2023efb2} further reveals that many EFBs connect via unsecured public Wi-Fi, LTE, or hotel networks without adequate authentication or isolation, exposing them to malware infection or unauthorized access. More critically, \cite{pentestpartners2023efb3a} demonstrate that tampering with EFB-based performance-calculation applications, such as altering runway length, surface condition, or aircraft weight, can result in incorrect takeoff parameters, potentially leading to runway excursions. Even when pilots perform verification, airlines that rely on a single EFB device for takeoff computation remain vulnerable to unnoticed manipulation. Finally, \cite{pentestpartners2023efb3b} show that critical takeoff and configuration data stored locally on EFB databases can be silently modified, underscoring the risk of data integrity compromise within an ecosystem increasingly built on heterogeneous hardware and non-standardized software. Figure \ref{fig:EFB} shows how different types of target data in the EFB can be exposed to various attack vectors.

\paragraph{Defense Mechanisms:}

To address the growing attack surface of connected EFB architectures, recent studies propose both proactive and layered defense approaches. \cite{bitton2019machine} developed a network-based Intrusion Detection System (NIDS) that employs machine learning and anomaly detection to secure Remote Desktop Protocol (RDP) and Remote Framebuffer Protocol (RFB) connections between commercial tablets and EFB servers. The system classifies Transmission Control Protocol (TCP) traffic at the packet level using a decision tree and k-means clustering, detecting malicious packets carrying real exploits with a true positive rate of 0.863 and a false positive rate of 0.0001. This fine-grained detection mechanism functions as an inline proxy between the EFB client and server, preventing code execution, privilege escalation, and authentication bypass attacks without degrading operational performance. \cite{true2021cybersecurity} describes a defense-in-depth architecture that combines robust software design, firewalls between avionics domains, certificate-based authentication, application-level encryption, and link-layer integrity checks. The inclusion of segregated networks such as the Airline Information Services Domain (AISD), the Aircraft Control Domain (ACD), and the Passenger Information and Entertainment Services Domain (PIESD), along with controlled physical access and redundant data links, further enhances availability and resilience. Together, these mechanisms establish both algorithmic and architectural safeguards against EFB-originated or propagated cyber threats.

\paragraph{Gaps and Open Problems:}

Despite the growing awareness of EFB-related cybersecurity vulnerabilities, current research remains fragmented and predominantly focused on usability, human factors, or hardware reliability rather than comprehensive adversarial threat modeling. \cite{bhardwaj2019safety} emphasize that EFB certification and regulatory frameworks still center on operational and environmental compliance, such as decompression, battery safety, and usability, without mandating security-by-design practices. Although \cite{habler2023assessing} extend the MITRE ATT\&CK (Adversarial Tactics, Techniques, and Common Knowledge) and STRIDE (Spoofing, Tampering, Repudiation, Information Disclosure, Denial of Service (DoS), and Elevation of Privilege) frameworks to aviation systems, there is still no validated taxonomy tailored to EFB-specific threat vectors, such as tampering of performance-calculation applications or manipulation of local databases, as demonstrated in recent operational analyses \cite{pentestpartners2023efb3a,pentestpartners2023efb3b}. Moreover, while recent work proposes machine-learning-based intrusion detection for EFB server communications \cite{bitton2019machine} and defense-in-depth architectures for flight-deck data exchange \cite{true2021cybersecurity}, these studies are limited to controlled environments and do not evaluate end-to-end resilience under realistic airline or eVTOL operational conditions. There remains a lack of standardized datasets, validation frameworks, and cross-domain testing methodologies for assessing how EFB-originated threats could propagate into avionics networks. Consequently, the integration of EFB cybersecurity into system-level safety assessments, coupled with empirical evaluation of proposed defenses in distributed and networked cockpit architectures, represents a critical and unresolved research frontier.


\subsubsection{FMS and Avionics Software Supply Chain}

\begin{table}[t]
\centering
\scriptsize
\caption{FMS and avionics software supply-chain risks for eVTOL and AAM operations.}
\label{tab:fms_supplychain}
\begin{tabularx}{0.99\linewidth}{%
    >{\raggedright\arraybackslash}p{0.18\linewidth} |
    >{\raggedright\arraybackslash}p{0.26\linewidth} |
    >{\raggedright\arraybackslash}p{0.26\linewidth} |
    >{\raggedright\arraybackslash}X}
\toprule
\rowcolor{black!10}
\textbf{Vulnerability} &
\textbf{Failure mode} &
\textbf{Impact on FMS} &
\textbf{Mitigation} \\
\midrule

GNSS disruption feeding the FMS &
Jamming or spoofing corrupts GNSS-derived position/velocity estimates \cite{truffer2017jamming,geister2018impact}. &
Incorrect aircraft state estimates driving wrong lateral/vertical path computations, unstable approaches, and incorrect leg transitions. &
Cross-check GNSS with inertial and barometric sources; anomaly recognition procedures; authenticated ACARS links \cite{geister2018impact}. \\[0.25em]

Navigation database tampering &
Modified or corrupted approach/runway records \cite{geister2018impact,habler2023assessing}. &
Incorrect glide paths, minima, missed-approach points, or procedure legs → potential CFIT or corridor deviations. &
Controlled DB processes, cryptographic integrity checks, AISD/ACD isolation \cite{habler2023assessing,shetty2008system}. \\[0.25em]

Inter-domain connectivity (ARINC 664) &
Ethernet-based avionics enlarge cross-domain attack surfaces \cite{habler2023assessing}. &
Injected or delayed messages entering the FMS route manager or sensor integration modules → corrupted internal state or stale routing updates. &
Segmentation, encrypted cross-domain pathways, ARINC-664-aware intrusion detection \cite{8260283,howard2006security,habler2023assessing}. \\[0.25em]

Software supply-chain compromise &
Malicious upstream components propagate into FMS/autonomy software \cite{Okafor2024,Martinez2021}. &
Incorrect route generation, hidden logic triggers, degraded guidance algorithms, or inconsistent performance computations. &
Supply-chain hardening, SBOM transparency, version attestation \cite{Welsh2025,Dong2025}. \\[0.25em]

Human and organizational factors &
Operators patch issues informally, masking underlying problems \cite{no2025study}. &
Incorrect FMS entries, inconsistent performance profiles, or unreported anomalies leading to degraded trajectory planning. &
Safety-II practices, structured reporting, collaborative monitoring \cite{no2025study}. \\
\bottomrule
\end{tabularx}
\end{table}

The FMS is an integrated suite designed to achieve optimal flight control through the coordination of multiple aircraft subsystems. The FMS enables automation across all phases of flight and provides flight crews with the necessary operational data through a unified interface. Its primary component, the Flight Management and Guidance Computer (FMC), computes the aircraft’s three-dimensional position, performance metrics, and other critical parameters required for precise and efficient flight along a predefined trajectory. These computations are derived from both manually entered and automatically acquired data. The Multipurpose Control and Display Unit (MCDU) serves as the pilot’s interface for data entry and system communication with the FMC. The Flight Control Unit (FCU) governs the aircraft’s lateral and vertical flight profiles, while the Flight Management Source Selector manages the selection of input sources used by the FMC. Finally, the display system presents key flight data and system information to the flight crew in real time \cite{ribaric2023aviation}.

\paragraph{Existing Threats:}
 
The FMS, as a core element of the aircraft control domain, is increasingly exposed to cyber threats due to its integration with open communication protocols and external data links such as GNSS and ACARS. \cite{truffer2017jamming} shows how  GNSS disruptions, even at modest jammer powers could corrupt the outputs of the FMS. Additionaly, simulation studies in \cite{geister2018impact} demonstrated that spoofing and jamming of GNSS signals can cause erroneous position inputs into the FMS during critical approach phases, while ACARS—used for flight plan and performance data uplink—lacks authentication and can be exploited to inject falsified route data or modify flight parameters. Moreover, they show that compromising the FMS navigation database itself poses a critical integrity threat: by altering stored approach data such as runway threshold altitudes or final approach fix geometry, attackers could cause the FMS to compute descent profiles leading below the intended glide path, potentially resulting in controlled flight into terrain under low-visibility conditions. As shown in \cite{habler2023assessing}, FMS is categorized within the Aircraft Control Domain (ACD), where interconnection between domains (ACD, AISD, and passenger domains) and the transition to Ethernet-based avionics (ARINC~664) expand the attack surface. Attack vectors include malicious data injection, cross-domain privilege escalation, and exploitation of unsecured wireless links. From a broader ATM perspective, \cite{liu2024optimal} emphasizes that vulnerabilities across surveillance (ADS-B), communication, and navigation subsystems can propagate to the FMS, as demonstrated by prior events where false ADS-B and ACARS data led to spurious guidance or control anomalies. They further highlight that the growing interconnection between airborne and ground infrastructures has transformed ATM into an open cyber–physical ecosystem, where attackers can exploit correlated vulnerabilities and, through reinforcement learning and structured threat graphs, autonomously identify and chain multiple weaknesses, potentially propagating cyberattacks from ATM subsystems into the FMS itself. These works converge on the finding that modern FMSs—once isolated—now operate within a networked cyber–physical ecosystem where inadequate segmentation \cite{shetty2008system}, legacy protocols, and lack of cryptographic protection represent systemic weaknesses.

Given that modern eVTOL FMS and autonomy modules will depend on software components, cloud-based routing services, and third-party data feeds, there is a real risk of software supply-chain compromise. Foundational research such as  \cite{Okafor2024}, demonstrates that supply chains lacking properties such as transparency, validity, and separation are vulnerable to poisoning and unauthorized modification. In addition, frameworks and methodologies such as \cite{Welsh2025}, \cite{Dong2025} provide taxonomies, checklists, and defense strategies for securing these chains in critical infrastructure. Real-world incidents, such as the SolarWinds breach, further illustrate how upstream compromise can propagate widely across dependent systems~\citep{Martinez2021}. Although these risks are general to any software supply chain, they indicate a potentially high risk for eVTOL applications. 

\paragraph{Defense Mechanisms:}

Across the reviewed studies, several layers of defense mechanisms are proposed to mitigate these vulnerabilities. \cite{geister2018impact} recommends procedural defenses such as enhanced pilot training for anomaly recognition and operational cross-checks between independent navigation sources to detect spoofed inputs. They also propose system-level mitigations including authentication of ACARS messages and the use of redundant inertial or barometric navigation references. \cite{habler2023assessing} extends these measures through a taxonomy of mitigation techniques based on the STRIDE threat model \cite{8260283, howard2006security}, including network segmentation between avionics domains, encryption of inter-domain data exchanges, and adoption of intrusion detection systems tailored for ARINC~664 traffic. Complementing these technical defenses, \cite{no2025study} stresses the human and organizational dimension, advocating Safety-II resilience models where crews and operators adaptively respond to system irregularities, transforming potential failure cases (e.g., FMS misbehavior) into resilient success through anticipatory monitoring and collaborative response protocols through effective communication between agents in the system.

\paragraph{Gaps and Open Problems:}

While taxonomies such as MITRE ATT\&CK \cite{strom2018mitre} have been extended to aviation, offering a structured framework to map multi-stage attack chains and support proactive vulnerability assessment, no unified or validated threat model yet exists specifically for avionics and FMS integration, leaving limited empirical evidence on real-world exploitability. \cite{geister2018impact} underscores the lack of large-scale experimental validation under operational conditions—existing simulator trials are limited in scope and primarily focus on pilot reactions rather than system resilience metrics. \cite{liu2024optimal} identifies insufficient automation in vulnerability correlation and attack-path discovery for complex ATM–FMS networks, calling for reinforcement-learning-driven knowledge graphs to anticipate evolving attack patterns. Finally, \cite{no2025study} reveals a socio-technical gap: organizational and training frameworks lag behind technological advances, with safety management systems remaining reactive rather than predictive. Collectively, these studies indicate that future research must integrate system-level cybersecurity modeling, real-time anomaly detection, and human-centered resilience frameworks to achieve trustworthy and adaptive FMS protection.


\subsection{Potential risks from aircraft automation, autonomy and connectivity}

The risks and vulnerabilities in this subsection are not common in commercial aviation aircraft, helicopters and general aviation aircraft yet, but they are expected with increasing automation and connectivity from the novel AAM aircraft, such as Joby, Wisk, Archer, EHang, etc.


\subsubsection{Cameras and Visual Sensors}

Visual sensors are a cornerstone of autonomous eVTOL operations because they provide rich situational awareness and centimeter-level precision without depending on external navigation infrastructure. In eVTOL missions, onboard cameras support critical perception tasks such as pad detection, obstacle avoidance, and precision landing, where radar or GPS alone may fail due to multipath interference, urban canyon effects, or satellite denial. The motivation for using visual sensing systems stems from their ability to emulate a pilot’s perception while enabling real-time semantic understanding of complex urban environments. These sensors, when fused with inertial or LiDAR data, allow eVTOLs to identify landing zones, estimate motion during GNSS outages, and achieve safe autonomy under constrained weight and energy budgets. NASA’s vision-based datasets demonstrate that camera-centric approaches can reconstruct reliable approach and landing trajectories even under variable illumination and cluttered rooftop environments \cite{brown2024visual}. Complementary research highlights that vision systems integrated with inertial and marker-based localization enable precise vertiport detection and multi-stage descent strategies that maintain sub-meter accuracy without GPS \cite{xiang2022multi}. Furthermore, comprehensive reviews emphasize that vision and perception subsystems are indispensable for eVTOL autonomy, forming the sensory foundation for navigation, safety assurance, and decision-making across all phases of flight \cite{xiang2024autonomous}. Table \ref{tab:sensing_perception_task_with_icons} demonstrates a mapping between sensing, perception, and the corresponding task. \cite{volocopter2023demoflight, skydrive2024vertiport} show real-world eVTOL operations at designated vertiports, illustrating how onboard visual sensing and scene understanding support safe approach and landing procedures in complex urban environments.

\newcolumntype{L}[1]{>{\raggedright\arraybackslash}p{#1}}

\begin{table*}[t]
    \centering
    \setlength{\tabcolsep}{6pt}
    \renewcommand{\arraystretch}{1.4}
    \begin{tabular}{L{0.30\textwidth} L{0.30\textwidth} L{0.30\textwidth}}
        \toprule
        \rowcolor{black!5}
        \textbf{Sensing} & \textbf{Perception} & \textbf{Task} \\
        \midrule
        \textbf{\faMapMarker*~GPS} – Provides absolute global position (latitude, longitude, altitude) used for georeferencing and navigation. 
        & \textbf{\faCogs~Visual–inertial odometry (VIO)} – Estimates precise relative motion by fusing camera and IMU data to track position and orientation when GPS is unreliable. 
        & \textbf{\faRoute~Path planning} – Generates safe, energy-efficient trajectories considering flight constraints and environmental conditions. \\[4pt]

        \textbf{\faLocationArrow~IMU} – Measures linear acceleration and angular velocity for estimating attitude, velocity, and short-term motion dynamics. 
        & \textbf{\faProjectDiagram~Semantic segmentation} – Classifies each pixel of an image into categories (e.g., sky, building, ground) to enable terrain and obstacle understanding. 
        & \textbf{\faPlaneArrival~Landing guidance} – Identifies safe landing zones and aligns vehicle trajectory for vertical descent using fused sensor data. \\[4pt]

        \textbf{\faVideo~Vision cameras} – Capture high-resolution imagery for scene perception, localization, and visual tracking. 
        & \textbf{\faBullseye~Object detection} – Recognizes and localizes key objects (e.g., vehicles, people, obstacles) within the visual scene for safe maneuvering. 
        & \textbf{\faExclamationTriangle~Obstacle avoidance} – Detects, predicts, and re-routes around obstacles in real-time using fused visual and range data. \\[4pt]

        \textbf{\faBroadcastTower~Ultrasonic, LiDAR, radar, or range sensors} – Provide depth, distance, and velocity information for obstacle proximity and 3D environment reconstruction. 
        & \textbf{\faGlobeAmericas~Scene understanding} – Builds a high-level situational model of the environment by integrating semantic, geometric, and dynamic cues. 
        & \textbf{\faTasks~Specified mission tasks} – Executes higher-level autonomous functions (e.g., inspection, delivery, surveillance) based on mission goals. \\
        \bottomrule
    \end{tabular}
    \caption{Functional mapping of sensing, perception, and task modules in eVTOL and autonomous aerial systems.}
    \label{tab:sensing_perception_task_with_icons}
\end{table*}

\paragraph{Existing Threats:}

NASA researchers formalized sensing and cybersecurity considerations for AAM vertiports, emphasizing that glass reflections, signage, and bright illumination can distort camera perception and compromise pad-detection accuracy \cite{mendonca2022advanced}. Further studies quantified the impact of adverse lighting, rain, and specular reflections on VTOL vision-aided navigation and recommended multi-sensor redundancy to mitigate environmental degradation \cite{veneruso2022sensing}. At the algorithmic level, physical adversarial patches and projected light patterns can mislead deep-learning detectors, showing that aerial vision models are susceptible to optical spoofing \cite{eykholt2018robust}. Universal adversarial patches capable of inducing false classifications across viewpoints have also been demonstrated \cite{brown2017adversarial}, while vulnerabilities in multi-sensor fusion indicate that simultaneous camera–LiDAR attacks can corrupt integrated perception pipelines \cite{cao2021invisible}. Even without adversaries, common corruptions such as blur, noise, and brightness shifts drastically degrade model accuracy, underscoring the brittleness of current vision models in real-world conditions \cite{hendrycks2019robustness}. Collectively, these studies establish that eVTOL vision systems are vulnerable to both natural interference and intentional optical deception. \cite{lu2021scale} introduced a scale-adaptive adversarial patch framework, called Patch-Noobj, which dynamically adjusts the patch size according to the aircraft’s scale in the image to make the aircraft vanish from detection results. Unlike pixel-level digital noise, these physically printable patches can mislead detectors such as YOLOv3, YOLOv5, and Faster R-CNN, reducing the average precision by up to 48 percent across multiple datasets. This attack demonstrates that eVTOL vision modules relying solely on CNN-based object detection could be compromised by small, visually innocuous overlays. They also illustrate how clean aircraft detections are removed after applying such adversarial patches, underscoring a realistic threat to safety-critical perception systems.

\begin{figure*}[h!]
    \centering
    \begin{minipage}[b]{0.48\textwidth}
        \centering
        \includegraphics[width=\textwidth]{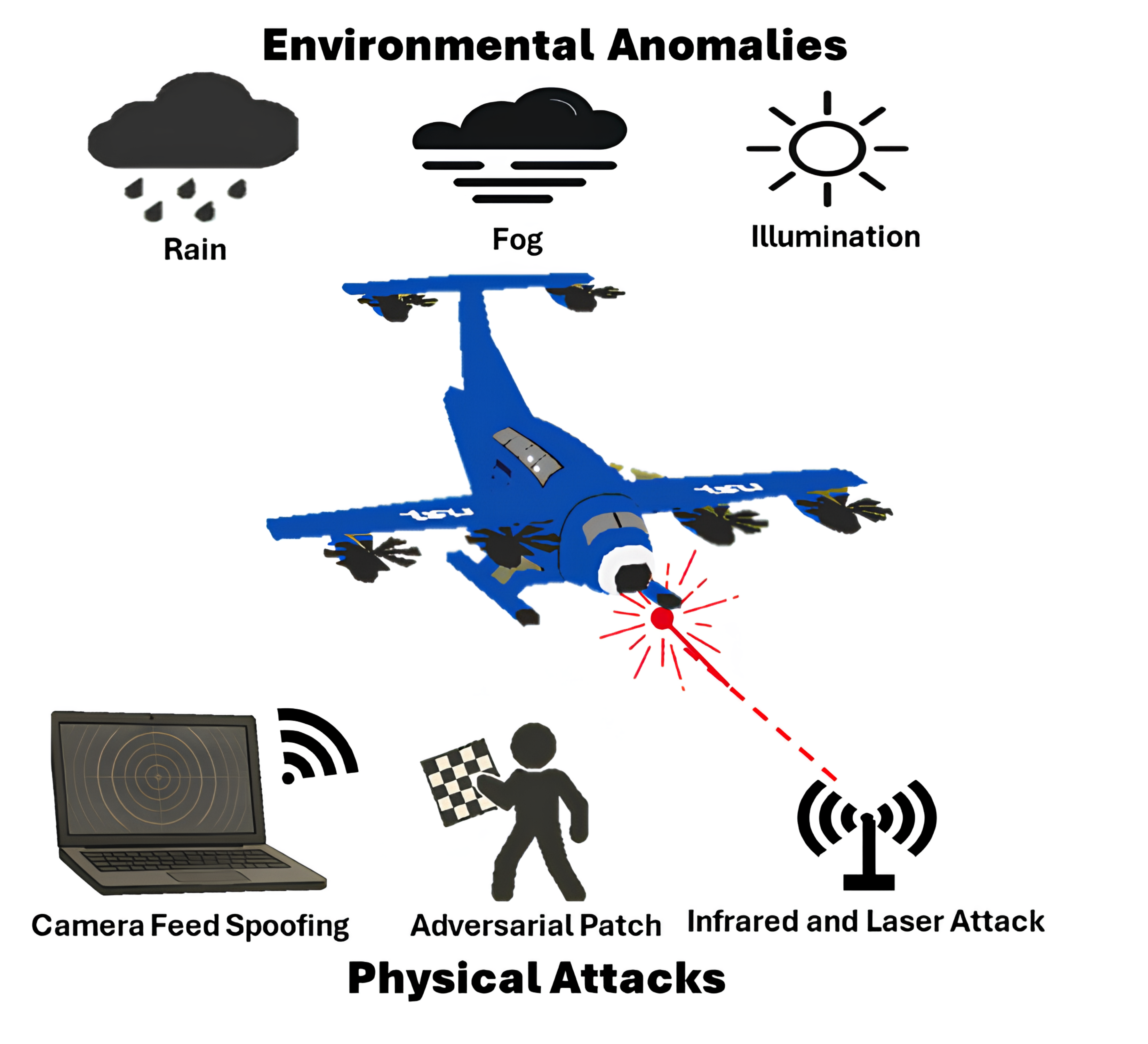}
        \caption{Threat channels affecting eVTOL visual perception on approach.}
        \label{fig:threat_channels}
    \end{minipage}
    \hfill
    \end{figure*}


\paragraph{Defense Mechanisms:}

To counter these vulnerabilities, several researchers have advanced system-level and model-level defenses. A LiDAR–Camera cross-verification framework was proposed to detect spoofing through geometric consistency checks \cite{zhang2023cooperative}. The Vision-Based Distributed Sensing (VIDIS) architecture integrates multiple cameras and inertial units to maintain pad visibility under illumination changes \cite{kawamura2025vision}, while complementary work extended this into a distributed vertiport sensor network fusing radar, IMU, and camera data for integrity monitoring \cite{ippolito2025enabling}. Explainable-AI-driven, certifiable fusion designs suitable for Size, Weight, and Power (SWaP) constrained avionics have been explored \cite{mishra2023autonomous, hu2025development}. Corruption-aware training via AugMix improves baseline robustness \cite{hendrycks2020augmix}, and out-of-distribution screening through Outlier Exposure and energy-based anomaly detection provides complementary safety layers \cite{hendrycks2019oe, liu2020energy}. To defend against localized tampering, PatchGuard \cite{xiang2021patchguard} and PatchCleanser \cite{xiang2022patchcleanser} restrict receptive-field influence and provide provable robustness guarantees. Addressing supply-chain threats, Neural Cleanse identifies and mitigates backdoor attacks \cite{wang2019neural}, while STRIP offers runtime trojan detection \cite{gao2019strip}. Together, these contributions outline a layered perception-security paradigm encompassing cross-sensor verification, robust model training, anomaly detection, and integrity auditing.

\paragraph{Gaps and Open Problems:}

Despite this progress, open challenges persist. No standardized AAM benchmarks currently evaluate perception under combined environmental, adversarial, and multi-modal perturbations. Certification procedures for machine-learned defenses remain undefined \cite{mishra2023autonomous, hu2025development}, while regulatory frameworks such as FAA and EUROCAE focus on lighting and geometry but omit adversarial validation criteria. The field therefore lacks eVTOL-specific datasets and evaluation protocols that integrate weather, motion, and spoofing factors. Moving forward, researchers and regulators must jointly establish explainable and certifiable fusion frameworks supported by benchmark-driven robustness metrics and dynamic trust scoring to ensure safe, verifiable visual perception in Advanced Air Mobility operations \cite{levitt2023uam}.


\subsubsection{Command/Control (C2) Link Attacks}

\begin{table}[t]
    \centering
    \scriptsize
    \caption{Command and control (C2) link threats across heterogeneous AAM networks.}
    \label{tab:c2_link_security}
    \begin{tabular}{|p{0.12\linewidth}|p{0.32\linewidth}|p{0.44\linewidth}|}
        \hline
        \rowcolor{black!10}
        \textbf{Threat family} & \textbf{Scenario} & \textbf{Mitigation strategies} \\
        \hline

        \multirow{2}{*}{Availability}
        & Selective jamming of 5G or A2G links during handovers disrupts control loops and triggers emergency loiter or diversion behaviors \cite{tang2021uamcyber,stouffer2020cns,kuba2024tcas}. 
        & Dual-path diversity (for example 5G plus 5030–5091 MHz aviation channels) and spread-spectrum or beamforming techniques for robust links \cite{faa2016uasC2,ullah2025cns5g,smith2020cockpit}. \\ \cline{2-3}

        & Frequent handovers and roaming at low altitude introduce timing vulnerabilities and gaps \cite{stouffer2020cns}. 
        & Empirical characterization of handover behavior and robust fallback modes that maintain safe loiter or return-to-home behavior. \\ \hline

        \multirow{2}{*}{Integrity}
        & Adversary manipulates or delays control commands by exploiting weak authentication or freshness guarantees in 5G or A2G protocols \cite{fonyi2024_5gsec,mitre2023uam5g}. 
        & Authenticated encryption with AEAD, DTLS or TLS with secure time sources, and strong anti-replay windows \cite{faa2024acas,rescorla2018transport}. \\ \cline{2-3}

        & Misconfigured network slices or roaming policies introduce unexpected latency spikes or expose C2 traffic to untrusted domains \cite{stouffer2020cns,mitre2023uam5g}. 
        & Slice binding and policy-controlled roaming that cryptographically ties credentials to authorized slices and operators \cite{strohmeier2020airground,ukwandu2022cyberaviation}. \\ 
        \hline
    \end{tabular}
\end{table}


\paragraph{Existing Threats:}
The command-and-control (C2) link in UAM operations is a crucial channel that transmits important information from the ground control to an automated aircraft (e.g. Wisk's aircraft), such as flight plan updates, contingency commands, etc. These C2 links may incorporate diverse communication infrastructures, including cellular networks (4G/5G), aviation-specific air-to-ground (A2G) channels, and satellite communication (SATCOM) systems \cite{tang2021uamcyber,stouffer2020cns}. This multi-link configuration enhances connectivity but expands the cyber attack surface, as any active bearer can become a target for jamming, spoofing, or session hijacking.

A primary threat concerns availability attacks through continuous or reactive jamming. Selective interference, especially during handovers between cellular and aviation links, can interrupt control loops or trigger automated autopilot fallbacks such as loiter, divert, or return-to-home maneuvers \cite{kuba2024tcas}. Even though the Federal Aviation Administration (FAA) and International Telecommunication Union (ITU) have reserved the 5030–5091~MHz band for AAM C2 operations \cite{faa2016uasC2}, many early UAM concepts depend on commercial or unlicensed spectrum for redundancy and cost efficiency \cite{ullah2025cns5g}. An adversary equipped with modest radio-frequency (RF) capability can exploit these channels by raising the noise floor or selectively jamming uplink control packets. In addition, in low altitudes, the aircraft experience frequent handovers and roaming across network providers. In addition, brief synchronization loss or misconfigured 5G network slices can introduce unacceptable latency or control gaps for UAM systems \cite{stouffer2020cns}.

Integrity and confidentiality attacks are also equally significant. Weak freshness guarantees or missing authentication in 5G or A2G protocols invite replay, man-in-the-middle (MitM), or impersonation attacks \cite{fonyi2024_5gsec,mitre2023uam5g}. A compromised ground gateway could inject delayed or falsified commands, while an adversary positioned between network nodes might suppress critical alerts or modify telemetry streams. These vulnerabilities underscore the absence of a unified, aviation-grade security baseline for C2 link.

\paragraph{Defense Mechanisms:}
Defending the C2 channel requires coordinated safeguards across the radio, network, and protocol layers. At the physical layer, radio-link robustness can be enhanced through spread-spectrum signaling, adaptive beamforming or null-steering, and dual-path diversity (e.g., 5G + dedicated A2G), as recommended in recent AAM communications resilience studies \cite{smith2020cockpit}. Such diversity enables graceful degradation rather than total control loss during interference. At the data-link and transport layers, authenticated encryption with anti-replay protection should be mandatory. Datagram Transport Layer Security (DTLS) or Transport Layer Security (TLS) using Authenticated Encryption with Associated Data (AEAD), coupled with secure time sources, can ensure both confidentiality and freshness in control messages \cite{faa2024acas}. In multi-operator networks, enforcing slice binding and policy-controlled roaming helps preserve end-to-end trust. By cryptographically binding each credential to an authorized network slice and disallowing roaming beyond approved domains, the system can prevent cross-operator spoofing and hijacking attacks \cite{strohmeier2020airground}. Together, these measures provide layered assurance against interference, replay, and unauthorized control insertion.

\paragraph{Gaps and Open Problems:}
Despite ongoing standardization efforts, there is limited vendor-neutral data on handover reliability for low-altitude, high-mobility eVTOL operations. Public datasets covering link transitions at urban altitudes and speeds remain scarce, making it difficult to certify acceptable fallback or recovery behaviors across heterogeneous network providers \cite{ukwandu2022cyberaviation}. Additionally, there is no harmonized definition of certifiable C2 fallback behavior across commercial, aviation, and satellite operators. Future work should prioritize empirical studies on link-layer resilience at eVTOL altitudes, formal definitions of fail-safe C2 states, and development of certifiable standards for secure handover and slice-based policy enforcement within the UAM communications architecture.

\subsubsection{Telemetry Data Link Attacks}

\begin{table}[t]
    \centering
    \scriptsize
    \renewcommand{\arraystretch}{1.2}
    \caption{Telemetry data links for eVTOL fleets: potential risks and early defense concepts.}
    \label{tab:telemetry_links}
    \begin{tabularx}{\linewidth}{>{\columncolor{RowGray}}p{0.32\linewidth} X}
        \toprule
        \textbf{Concern} & \textbf{Implication and mitigation } \\
        \midrule

        Dependence on heterogeneous networks
        & Telemetry often traverses airport networks, public clouds, and vendor platforms, creating a broad attack surface \cite{Rieder2023AAMSecurity,skydrive2024vertiport}. 
          Network segmentation and system hardening are recommended to isolate telemetry paths from general IT infrastructure \cite{Rieder2023AAMSecurity}. \\

        Indirect effects of avionics bus attacks
        & ARINC 429 flooding can suppress or delay navigation or system-status data which may feed ground telemetry streams \cite{trask2024arinc}. 
          Bus monitoring, rate limiting, and redundant channels can help preserve critical data under denial-of-service conditions \cite{trask2024arinc}. \\

        Lack of demonstrated telemetry-specific attacks
        & Existing work highlights telemetry as a potential target but provides no experimental exploitation of eVTOL links \cite{Rieder2023AAMSecurity}. 
          Encrypted telemetry with authentication and anti-replay, along with anomaly detection based on baseline cross-checking, offers a forward path \cite{USP20200013243A1}. \\
        \bottomrule
    \end{tabularx}
\end{table}


\paragraph{Existing Threats:}
Telemetry data links are communication channels that transmit real-time aircraft information, including position, speed, system status, navigation data, aircraft health status, battery information, and other sensor readings, between the aircraft and ground-based systems such as flight monitoring centers, air traffic management, or fleet operations platforms~\cite{Rieder2023AAMSecurity}. In eVTOL and urban air mobility (UAM) operations, telemetry links are critical for enabling autonomous flight management, collision avoidance, traffic coordination, and integration with airport networks and AAM ecosystems. Current literature does not report experimentally demonstrated telemetry data link attacks specifically targeting eVTOLs. The ARINC~429 study~\cite{trask2024arinc} shows that bus-flooding denial-of-service attacks can suppress navigation data in conventional avionics systems, which could indirectly affect telemetry streams if  data is transmitted to ground operators. SkyGrid~\cite{skydrive2024vertiport} and the PARAS report~\cite{Rieder2023AAMSecurity} highlight that telemetry links, particularly those dependent on cloud-based services or airport networks, represent potential cyber-attack surfaces. 
However, no actual attacks or incidents are documented, making this threat largely hypothetical.
\paragraph{Defense Mechanisms:}
While direct telemetry attacks on eVTOLs have not been demonstrated, several defensive strategies are suggested in the literature. SkyGrid~\cite{skydrive2024vertiport} emphasizes the use of secure communication frameworks, including encryption, authentication, and resilient cloud-based protocols, to protect the integrity and confidentiality of telemetry data. Also \cite{Rieder2023AAMSecurity} recommends network segmentation and system hardening to isolate telemetry networks from broader airport IT systems and to enable monitoring for anomalous traffic. ARINC~429~\cite{trask2024arinc} implies that redundancy, rate limiting, and bus-monitoring mechanisms could help mitigate potential denial-of-service disruptions that affect telemetry. Additionally, techniques from US Patent~\cite{USP20200013243A1}, such as cross-checking incoming data against trusted baselines, could be adapted to monitor telemetry streams for anomalies, although these methods have not been specifically validated in eVTOL operations.
\paragraph{Gaps and Open Problems:}
There is no unified standard or certification guidance for secure telemetry operations or onboard anomaly detection in eVTOLs. The end-to-end telemetry chain in eVTOLs involves heterogeneous networks, including onboard avionics buses, wireless communication links, and cloud services, and the literature notes that protecting this integrated system remains an open challenge~\cite{Rieder2023AAMSecurity, skydrive2024vertiport}. Finally, proposed defense techniques such as encryption, redundancy, or baseline verification have not been validated in operational eVTOL systems or tested against realistic threat scenarios. Overall, telemetry links are recognized as critical for eVTOL operations, yet both threat evaluation and standardized defense mechanisms remain largely unexplored areas for future research in urban air mobility (UAM) cybersecurity.

\subsection{Dataset and data stream attacks}

\subsubsection{Attacks on the Navigation Database (Data Attacks)}

\begin{table}[t]
    \centering
    \scriptsize
    \caption{Navigation database risks for low-altitude eVTOL operations.}
    \label{tab:navdb_risks}
    \begin{tabularx}{0.98\linewidth}{%
        >{\raggedright\arraybackslash}p{0.22\linewidth} 
        >{\raggedright\arraybackslash}p{0.28\linewidth}
        X}
    \toprule
    \rowcolor{black!10}
    \textbf{Element} &
    \textbf{Risk / Impact} &
    \textbf{Mitigation} \\
    \midrule

    Vertiport and corridor records &
    Incorrect pad coordinates or corridor definitions may route aircraft into obstacles or restricted zones. &
    Provider-side cybersecurity, provenance tracking, and validation of nav-data layers \cite{JeppesenNavData,LidoDeveloperPortal2024,NavblueNavigationPlus,skydrive2024vertiport}. \\[0.3em]

    3D obstacle and terrain models &
    Manipulated obstacle/terrain data can hide structures or degrade separation margins. &
    Strong cloud-side data handling, airport network protections, and onboard plausibility checks \cite{Rieder2023AAMSecurity,skydrive2024vertiport}. \\[0.3em]

    Onboard nav-database handling &
    Corrupted or incomplete updates cause inconsistent or unsafe procedures. &
    Integrity checks comparing new updates to trusted baselines; automatic rejection and rollback \cite{USP20200013243A1}. \\[0.3em]

    Data-bus transport (ARINC 429) &
    Bus-level DoS delays or blocks internal nav-data transfers. &
    Bus health monitoring, rate limiting, and redundant avionics networks tailored for eVTOL \cite{trask2024arinc}. \\

    \bottomrule
    \end{tabularx}
\end{table}


\paragraph{Existing Threats:}
The eVTOL navigation database is a specialized digital dataset designed to support safe, automated, and efficient flight in UAM environments. It incorporates traditional aviation data such as waypoints, airspace boundaries, airports, navaids, and standard procedures, while also integrating UAM-specific elements, including vertiport locations, designated sky corridors, detailed 3D obstacle maps (buildings, power lines, bridges, cranes), geofenced no-fly zones, micro-weather layers, and energy-based routing information for electric propulsion. Because eVTOLs operate at low altitudes in dense urban areas, these databases must include high-resolution terrain and obstacle models, dynamic operational restrictions, and real-time updates. Major providers adapting their datasets for eVTOL operations include Jeppesen~\cite{JeppesenNavData}, Lido Navigation~\cite{LidoDeveloperPortal2024}, and Navblue~\cite{NavblueNavigationPlus}, alongside emerging UAM data and traffic management suppliers such as Altitude Angel, SkyGrid, and ANRA Technologies. Despite the rapid growth of eVTOL and UAM systems, there is a notable lack of publicly available reports or academic studies demonstrating attacks specifically targeting navigation databases for these aircraft.  Nevertheless, some sources, such as US Patent~\cite{USP20200013243A1}, describe methods for detecting and mitigating tampering of flight management system (FMS) navigation databases. Although the patent is not eVTOL-specific, the underlying technology—digital navigation databases, waypoints, and terrain/obstacle data—is also utilized in eVTOL systems, suggesting that database manipulation or corruption attacks could plausibly occur in this context. Similarly, while the ARINC 429 cyber-vulnerability study does not address eVTOLs directly, many eVTOL designs employ ARINC 429 for avionics communications. This implies that bus-flooding denial-of-service attacks, as demonstrated in the study\cite{trask2024arinc}, could technically affect eVTOL systems relying on the interface. 

\paragraph{Defense Mechanisms:}
Several sources highlight the need for
stronger cybersecurity practices, but few offer concrete, technically validated defenses
for protecting onboard navigation databases. SkyGrid's white paper emphasizes the importance of \textit{modern, cloud-centric cybersecurity frameworks} to protect the third-party navigation, routing, and UTM services on which eVTOLs depend~\cite{skydrive2024vertiport}. Their analysis identifies cloud service providers as high-value targets and stresses the need for resilient authentication, continuous monitoring, and strict data-handling controls to prevent manipulation of navigation or routing data delivered to aircraft. Similarly, \cite{Rieder2023AAMSecurity} warn that the increasing integration of eVTOLs with airport IT networks expands the cyber attack surface and requires improved network segmentation, data governance, and system hardening to protect critical operational data flows. The ARINC~429 denial-of-service study demonstrates the feasibility
of bus-flooding attacks that can suppress or delay navigation information on conventional
aircraft data buses~\cite{trask2024arinc}. While not developed for eVTOL platforms, this
work implies that eVTOL architectures relying on ARINC~429 may require rate limiting,
bus health monitoring, and redundant communication channels to mitigate similar
disruptions. The US patent,~\cite{USP20200013243A1}, on navigation
database tampering proposes onboard integrity checks that compare incoming aeronautical
data with trusted baselines to detect corruption or unauthorized modification of FMS
navigation datasets. These techniques represent promising building blocks
for eVTOL navigation-data protection, but lack specific validation or certification
guidance.


\paragraph{Gaps and Open Problems}
A gap in current research and industry practice is the absence of a unified, trusted provenance and integrity-assurance standard for eVTOL navigation data. While existing providers such as Jeppesen~\cite{JeppesenNavData}, Lido~\cite{LidoDeveloperPortal2024}, and Navblue~\cite{NavblueNavigationPlus} supply high-quality aeronautical datasets, there is limited publicly available guidance on guaranteeing the authenticity, lineage, and tamper resistance of the expanded UAM-specific data layers they now include (e.g., 3D city models, sky corridors, geofencing constraints).  A related open problem is the lack of certification requirements or technical standards for onboard auto-rejection logic capable of detecting and rejecting corrupted, manipulated, or inconsistent navigation database entries. 

\subsection{Cloud security concerns with AAM architecture and service providers}

\subsubsection{AAM Architecture and eVTOL--Cloud Interactions (PSUs, SDSPs, Provider APIs)}

\begin{table}[t]
    \centering
    \scriptsize
    \caption{Cloud-centric AAM ecosystem: risks at the PSU, SDSP, and API boundary.}
    \label{tab:cloud_aam_risks}
    \begin{tabularx}{0.95\linewidth}{%
        >{\raggedright\arraybackslash}p{0.20\linewidth} 
        X 
        X}
    \toprule
    \rowcolor{black!10}
    \textbf{Actor / interface} &
    \textbf{Role in AAM architecture \& cybersecurity concern} &
    \textbf{Mitigation and references} \\
    \midrule

    Provider of Services for UAM (PSU) &
    Manages flight intents, approval, and strategic deconfliction for eVTOL operations \cite{faaUAM2023,lewis2025developing}.  
    Compromised credentials or replayed tokens allow flight-plan manipulation or large-scale service disruption \cite{freeman2022immutable,Samanani_2025}. &
    Strong authentication and encryption (OAuth2 authorization, mutual TLS, and cryptographically signed payloads with bounded freshness) ensure that only verified entities exchange valid, non-replayable data \cite{campbell2020oauth,rescorla2018transport}.  
    \\[0.3em]

    Supplemental Data Service Providers (SDSPs) &
    Supply weather, traffic, terrain, and constraint layers consumed by PSUs and operators \cite{faaUAM2023}.  
    Poisoned or stale data can mislead routing and conflict detection across fleets \cite{freeman2022immutable,zhong2025enhancing}. &
    Redundancy and cross-validation among SDSPs and PSUs improve integrity by comparing rapidly changing inputs to flag inconsistencies \cite{zhong2025enhancing}.  
    If mismatches occur, graceful degradation reverts to the last known good state.  
    \\[0.3em]

    Provider APIs and shared cloud infrastructure &
    Provider APIs connect aircraft, operators, PSUs, SDSPs, and vertiports \cite{lewis2025developing}.  
    Insecure or deprecated API calls resemble connected-EV vulnerabilities, enabling data exfiltration or control overreach \cite{saleem2025cybersecurity}.  
    Cloud coupling can propagate cascading failures across operational domains \cite{freeman2022immutable}. &
    Secure API gateways, unified authentication/authorization, and immutable audit logs for forensic readiness \cite{freeman2022immutable,zhong2025enhancing}.  
    Long-term challenges include certifiable, scalable cloud architectures integrating provenance tracking and redundancy-aware validation.  
    \\

    \bottomrule
    \end{tabularx}
\end{table}

AAM systems, particularly those involving eVTOL aircraft, operate within a distributed digital ecosystem that depends heavily on cloud-based infrastructure. These systems support mission-critical functions such as flight plan approval, real-time conflict monitoring and alerting, weather forecast, and data exchange among diverse stakeholders. Unlike traditional aviation which has been dominated by centralized FAA control and fixed communication channels, the emerging AAM framework adopts a decentralized, service-oriented architecture \cite{faaUAM2023}. 

Within this model, several cloud-dependent entities play key operational roles. Providers of Services for UAM (PSUs) manage flight plans and strategic deconfliction, ensuring coordinated airspace use. Supplemental Data Service Providers (SDSPs) supply vital information such as weather updates, traffic surveillance, and terrestrial obstacle data, often distributed via Discovery and Synchronization Services. Meanwhile, Provider APIs serve as the connective framework, facilitating secure communication between PSUs, SDSPs, eVTOL aircrafts, and ground systems \cite{lewis2025developing}. Each link in this interconnected chain—from an aircraft’s onboard system to PSU and SDSP cloud interfaces—represents a potential cybersecurity exposure that must be carefully managed.

\paragraph{Existing Threats: }
Interactions between eVTOL aircraft and third-party cloud services expand the potential attack surface, exposing AAM infrastructure to familiar but high-impact cybersecurity threats. These include man-in-the-middle exploits, credential theft, data replay, and spear-phishing attacks capable of manipulating or corrupting critical operational data \cite{freeman2022immutable}. For instance, compromised access credentials or replayed authentication token could allow attackers to infiltrate PSU networks, inject unauthorized commands, or alter flight data in real time \cite{lewis2025developing}. Such breaches could disrupt mission operations—grounding entire eVTOL fleets or preventing authorized flights from departure \cite{Samanani_2025}.  

Provider APIs, which enable cross-system communication, also present a well-documented vector for exploitation. Vulnerabilities common in other connected electric vehicle (EV) platforms, such as deprecated or poorly secured API calls, could similarly be leveraged in AAM contexts to exfiltrate sensitive operational data or compromise flight integrity \cite{saleem2025cybersecurity}. These risks highlight the urgent need for rigorous authentication, continuous monitoring, and standardized cybersecurity frameworks across all cloud interaction layers within the AAM ecosystem.

\paragraph{Defense Mechanisms:  }AAM architectures can mitigate these threats through layered, standards-based security protocols. Strong authentication and encryption—such as OAuth2 authorization, mutual Transport Layer Security, and cryptographically signed payloads with bounded freshness \cite{campbell2020oauth, rescorla2018transport}—help ensure that only verified entities exchange valid and non-replayable data. Redundancy and cross-validation among PSUs and SDSPs further enhance integrity by comparing time-sensitive inputs (e.g., weather, flight intents, and airspace constraints) to identify inconsistencies or stale updates before they propagate through the system \cite{zhong2025enhancing}. In cases of service disruption or data mismatch, graceful degradation mechanisms allow eVTOL management systems to revert to the last known good state, defer new commitments, and alert human supervisors—while maintaining immutable audit trails to support post-incident forensics and traceability.

\paragraph{Gaps and Open Problems: } The AAM ecosystem lacks a unified, certifiable framework for trusted data exchange across service providers. Current FAA specifications define functional APIs and message formats but do not prescribe verifiable provenance, redundancy validation, or freshness requirements for data originating from PSUs, SDSPs, or third-party clouds \cite{faaUAM2023}. This absence complicates certification of safety-critical cloud services and prevents end-to-end assurance that flight systems consume authentic, timely information. 

Moreover, as illustrated in Figure 1, cloud connectivity through PSUs and SDSPs forms the critical information backbone linking UAM vehicles, operators, vertiports, and regulatory oversight. Compromised cloud channels can therefore amplify vulnerabilities across multiple operational domains simultaneously, yet no framework currently exists for reasoning about cascading failures or cross-system attack propagation through shared cloud infrastructure. Developing scalable, certifiable architectures that integrate provenance tracking, redundancy verification, cross-system integrity checks, and latency-aware security remains an open challenge for safe AAM deployment at scale.

\keptcontentend

\section{Conclusion} \label{sec: Conclusion}
The expansion of AAM and eVTOL technologies represents a major step toward more automated, connected, and digitally dependent air transportation. As these systems evolve, their safety and reliability will increasingly depend on how well emerging cyber-physical risks are understood and addressed. This paper examined the primary vulnerabilities affecting modern aerial platforms and outlined the technical and procedural measures that can strengthen their resilience.

Looking ahead, several areas demand urgent attention. First, the continued reliance on unauthenticated surveillance channels such as Mode~S, SSR, and ADS-B remains a critical exposure for collision avoidance and traffic coordination. Second, GNSS jamming and spoofing present a high-impact threat for eVTOL aircraft navigation in dense urban environments, where loss of positional integrity can rapidly lead to unsafe flight states. Third, the radio based voice communication, and communication channels for air-to-ground and air-to-air need to be secured. Fourth, the increasing dependence on camera-based perception and machine learning introduces new risks from adversarial manipulation, sensor degradation, and multi-modal spoofing. Fifth, cloud-connected AAM services, including PSUs and SDSPs, create systemic vulnerabilities that may propagate across fleets if provider APIs or synchronization services are compromised. Finally, electronic flight bags and onboard software supply chains require stronger guarantees of data provenance, update integrity, and device hardening to prevent operational misuse.

Addressing these risks will require close collaboration across avionic engineering, cybersecurity, flight operations, and regulatory communities, along with a commitment to integrating security and integrity safeguards into every stage of AAM system development, certification, and flight operation. As eVTOLs transition into everyday transportation systems, establishing trust in their digital resilience will be central to ensuring the safe and scalable deployment of future urban air mobility.

\section*{Acknowledgments}
This material is based upon work supported by the NASA Aeronautics Research Mission Directorate (ARMD) University Leadership Initiative (ULI) under cooperative agreement number 80NSSC24M0070. Any opinions, findings, and conclusions or recommendations expressed in this material are those of the author(s) and do not necessarily reflect the views of the National Aeronautics and Space Administration.

\bibliography{sample}

@article{zhou2023improvement,
  title={Improvement of GPS displacement measurement accuracy for high-rise buildings by machine learning},
  author={Zhou, Qi and Li, Qiu-Sheng and Han, Xu-Liang and Lu, Bin and Wan, Jun-Wen and Xu, Kang},
  journal={Journal of Building Engineering},
  volume={78},
  pages={107581},
  year={2023},
  publisher={Elsevier}
}

@article{brossard2020ai,
  title={AI-IMU dead-reckoning},
  author={Brossard, Martin and Barrau, Axel and Bonnabel, Silv{\`e}re},
  journal={IEEE Transactions on Intelligent Vehicles},
  volume={5},
  number={4},
  pages={585--595},
  year={2020},
  publisher={IEEE}
}

@article{um2020configuring,
  title={Configuring RTK-GPS architecture for system redundancy in multi-drone operations},
  author={Um, Inseop and Park, Seongjoon and Kim, Hyeong Tae and Kim, Hwangnam},
  journal={IEEE Access},
  volume={8},
  pages={76228--76242},
  year={2020},
  publisher={IEEE}
}

@inproceedings{rhudy2012fusion,
  title={Fusion of GPS and redundant IMU data for attitude estimation},
  author={Rhudy, Matthew and Gross, Jason and Gu, Yu and Napolitano, Marcello},
  booktitle={AIAA Guidance, Navigation, and Control Conference},
  pages={5030},
  year={2012}
}

@inproceedings{opromolla2016lidar,
  title={LIDAR-inertial integration for UAV localization and mapping in complex environments},
  author={Opromolla, Roberto and Fasano, Giancarmine and Rufino, Giancarlo and Grassi, Michele and Savvaris, Al},
  booktitle={2016 international conference on unmanned aircraft systems (ICUAS)},
  pages={649--656},
  year={2016},
  organization={IEEE}
}

@inproceedings{petrlik2021lidar,
  title={LiDAR-based stabilization, navigation and localization for UAVs operating in dark indoor environments},
  author={Petrl{\'\i}k, Mat{\v{e}}j and Krajn{\'\i}k, Tom{\'a}{\v{s}} and Saska, Martin},
  booktitle={2021 International Conference on Unmanned Aircraft Systems (ICUAS)},
  pages={243--251},
  year={2021},
  organization={IEEE}
}

@inproceedings{nguyen2016ultra,
  title={An ultra-wideband-based multi-UAV localization system in GPS-denied environments},
  author={Nguyen, Thien Minh and Zaini, Abdul Hanif and Guo, Kexin and Xie, Lihua},
  booktitle={2016 International Micro Air Vehicles Conference},
  volume={6},
  pages={1--15},
  year={2016}
}

@article{shule2020uwb,
  title={UWB-based localization for multi-UAV systems and collaborative heterogeneous multi-robot systems},
  author={Shule, Wang and Almansa, Carmen Mart{\'\i}nez and Queralta, Jorge Pe{\~n}a and Zou, Zhuo and Westerlund, Tomi},
  journal={Procedia Computer Science},
  volume={175},
  pages={357--364},
  year={2020},
  publisher={Elsevier}
}

@article{tang2020review,
  title={A Review on Cybersecurity Vulnerabilities for Urban Air Mobility},
  author={Anthony C. B. Tang},
  journal={NASA Secure Airspace},
  year={2020},
  note={NASA Ames Research Center}
}

@inproceedings{freeman2020cyber,
  title={A survey of cyber threats and security controls analysis for urban air mobility environments},
  author={Freeman, Kenneth and Garcia, Steve},
  booktitle={AIAA Scitech 2021 Forum},
  pages={0660},
  year={2021}
}

@techreport{faaUAM2023,
  title        = {Urban Air Mobility (UAM) Concept of Operations Version 2.0},
  author       = {{Federal Aviation Administration}},
  institution  = {U.S. Department of Transportation, Federal Aviation Administration},
  year         = {2023},
  month        = {April},
  number       = {Version 2.0},
  note         = {Available online: \url{https://www.faa.gov/sites/faa.gov/files/Urban%20Air%20Mobility%20%28UAM%29%20Concept%20of%20Operations%202.0_1.pdf}},
}

@ARTICLE{abuzaid2023communications,
  author={Zaid, Abdullah Abu and Belmekki, Baha Eddine Youcef and Alouini, Mohamed-Slim},
  journal={IEEE Communications Magazine}, 
  title={eVTOL Communications and Networking in UAM: Requirements, Key Enablers, and Challenges}, 
  year={2023},
  volume={61},
  number={8},
  pages={154-160},
  doi={10.1109/MCOM.004.2300061}}

@ARTICLE{wei2024autonav,
  author={Wei, Henglai and Lou, Baichuan and Zhang, Zezhong and Liang, Bohang and Wang, Fei-Yue and Lv, Chen},
  journal={IEEE Transactions on Intelligent Vehicles}, 
  title={Autonomous Navigation for eVTOL: Review and Future Perspectives}, 
  year={2024},
  volume={9},
  number={2},
  pages={4145-4171},
  doi={10.1109/TIV.2024.3352613}}

@INPROCEEDINGS{9338487,
  author={Ye, Siyuan and Wan, Zeyu and Zeng, Long and Li, Chenglong and Zhang, Yu},
  booktitle={2020 4th CAA International Conference on Vehicular Control and Intelligence (CVCI)}, 
  title={A vision-based navigation method for eVTOL final approach in urban air mobility(UAM)}, 
  year={2020},
  volume={},
  number={},
  pages={645-649},
  doi={10.1109/CVCI51460.2020.9338487}}

@techreport{maven2022vertiports,
  title = {Optimal Locations for Air Mobility Vertiports},
  author = {{Maven Research Inc.}},
  institution = {NASA Aeronautics Research Mission Directorate},
  year = {2022},
  note = {Contract NOIS2-049-RFTP},
  url = 
    {https://ntrs.nasa.gov/api/citations/20220005871/downloads/Final%20Report%20v2%20-%20ARMD%20Air%20Mobility%20Vertiports%20Maven%20Research%20Inc%2025JAN2022.pdf}
    }

@techreport{ram2021,
  title={Regional air mobility: Leveraging our national investments to energize the American travel experience},
  author={Antcliff, Kevin and Borer, Nicholas and Sartorius, Sky and Saleh, Pasha and Rose, Robert and Gariel, Maxime and Oldham, Joseph and Courtin, Christopher and Bradley, Marty and Roy, Satadru},
  institution={NASA},
  year={2021},
}

@inproceedings{chancey2023hat,
author = {Chancey, Eric and Politowicz, Michael and Buck, Bill and Ballard, Kathryn and Unverricht, James and Houston, Vincent and Saephan, Meghan and Le Vie, Lisa},
year = {2023},
month = {01},
pages = {},
title = {Foundational Human-Autonomy Teaming Research and Development in Scalable Remotely Operated Advanced Air Mobility Operations: Research Model and Initial Work},
doi = {10.2514/6.2023-1066}
}

@inproceedings{nasa2023distributed,
  title={Enabling Smart Urban Airspaces through Distributed Sensing Technologies},
  author={Ippolito, Corey A and Martin, Rodney A and Kawamura, Evan and Gorospe, George and Holforty, Wendy and Kannan, Keerthana and Stepanyan, Vahram and Lombaerts, Thomas and Brown, Nelson and Dolph, Chester},
  booktitle={AIAA SCITECH 2025 Forum},
  pages={0344},
  year={2025}
}

@inproceedings{eykholt2018robust,
  title={Robust physical-world attacks on deep learning visual classification},
  author={Eykholt, Kevin and Evtimov, Ivan and Fernandes, Earlence and Li, Bo and Rahmati, Amir and Xiao, Chaowei and Prakash, Atul and Kohno, Tadayoshi and Song, Dawn},
  booktitle={Proceedings of the IEEE conference on computer vision and pattern recognition},
  pages={1625--1634},
  year={2018}
}

@book{hofmann2012global,
  title={Global positioning system: theory and practice},
  author={Hofmann-Wellenhof, Bernhard and Lichtenegger, Herbert and Collins, James},
  year={2012},
  publisher={Springer Science \& Business Media}
}

@article{team2014global,
  title={Global positioning system (GPS) standard positioning service (SPS) performance analysis report},
  author={Team, GPS Product},
  journal={GPS Product Team: Washington, DC, USA},
  year={2014}
}

@article{boquet2024feasibility,
  title={Feasibility of Providing High-Precision GNSS Correction Data through Non-Terrestrial Networks},
  author={Boquet, Guillem and Vilajosana, Xavi and Martinez, Borja},
  journal={IEEE Transactions on Instrumentation and Measurement},
  year={2024},
  publisher={IEEE}
}

@article{chen2025obtpn,
  title={OBTPN: A vision-based network for UAV geo-localization in multi-altitude environments},
  author={Chen, Nanxing and Fan, Jiqi and Yuan, Jiayu and Zheng, Enhui},
  journal={Drones},
  volume={9},
  number={1},
  pages={33},
  year={2025},
  publisher={MDPI}
}

@inproceedings{kim2020precise,
  title={Precise localization of a UAV with single vision camera and deep learning},
  author={Kim, Hyeong Tae and Kim, Hwangnam},
  booktitle={GLOBECOM 2020-2020 IEEE Global Communications Conference},
  pages={1--6},
  year={2020},
  organization={IEEE}
}

@inproceedings{yol2014vision,
  title={Vision-based absolute localization for unmanned aerial vehicles},
  author={Yol, Aurelien and Delabarre, Bertrand and Dame, Amaury and Dartois, Jean-Emile and Marchand, Eric},
  booktitle={2014 IEEE/RSJ International Conference on Intelligent Robots and Systems},
  pages={3429--3434},
  year={2014},
  organization={IEEE}
}

@article{harman1989tcas,
  title={TCAS- A system for preventing midair collisions},
  author={Harman, William H},
  journal={The Lincoln Laboratory Journal},
  volume={2},
  number={3},
  pages={437--457},
  year={1989},
  publisher={Citeseer}
}

@article{kochenderfer2012next,
  title={Next-generation airborne collision avoidance system},
  author={Kochenderfer, Mykel J and Holland, Jessica E and Chryssanthacopoulos, James P},
  year={2012}
}

@techreport{Eurocontrol2017ACASGuide,
  author       = {{EUROCONTROL}},
  title        = {{ACAS Guide: Airborne Collision Avoidance, Chapter "Future of Collision Avoidance: ACAS X"}},
  institution  = {EUROCONTROL},
  year         = {2017},
  month        = {December},
  pages        = {21},
  note         = {Available from EUROCONTROL, Brussels, Belgium.},
  url          = {https://www.eurocontrol.int/sites/default/files/2019-03/safety-acas-2-guide.pdf}
}

@inproceedings{guendel2025collision,
  title={Collision Avoidance for Rotorcraft in Urban Airspace with ACAS Xr},
  author={Guendel, Randal E and Wu, Samuel},
  booktitle={AIAA AVIATION FORUM AND ASCEND 2025},
  pages={3668},
  year={2025}
}

@article{smith2022understanding,
  title={Understanding realistic attacks on airborne collision avoidance systems},
  author={Smith, Matthew and Strohmeier, Martin and Lenders, Vincent and Martinovic, Ivan},
  journal={Journal of transportation security},
  volume={15},
  number={1},
  pages={87--118},
  year={2022},
  publisher={Springer}
}

@article{habler2023assessing,
  title={Assessing aircraft security: A comprehensive survey and methodology for evaluation},
  author={Habler, Edan and Bitton, Ron and Shabtai, Asaf},
  journal={ACM Computing Surveys},
  volume={56},
  number={4},
  pages={1--40},
  year={2023},
  publisher={ACM New York, NY}
}

@inproceedings{kuba2024navigating,
  title={Navigating Threats: A Vulnerability Analysis of TCAS Interaction with Other Aircraft Systems},
  author={Kuba, Sakurako and Babiceanu, Radu F},
  booktitle={2024 AIAA DATC/IEEE 43rd Digital Avionics Systems Conference (DASC)},
  pages={1--6},
  year={2024},
  organization={IEEE}
}

@techreport{FAA2011TCASIntro,
  author       = {{Federal Aviation Administration}},
  title        = {{Introduction to TCAS~II Version~7.1}},
  institution  = {Federal Aviation Administration (FAA)},
  year         = {2011},
  url          = {https://www.faa.gov/documentlibrary/media/advisory_circular/tcas%20ii%20v7.1%20intro%20booklet.pdf}
}

@incollection{strohmeier2020securing,
  title={Securing the air--ground link in aviation},
  author={Strohmeier, Martin and Martinovic, Ivan and Lenders, Vincent},
  booktitle={The Security of Critical Infrastructures: Risk, Resilience and Defense},
  pages={131--154},
  year={2020},
  publisher={Springer}
}

@techreport{FAA_AC90_120,
  author       = {{Federal Aviation Administration}},
  title        = {{Advisory Circular AC 90-120: Operational Use of Airborne Collision Avoidance Systems}},
  institution  = {U.S. Department of Transportation, Federal Aviation Administration},
  year         = {2024},
  url          = {https://www.faa.gov/documentLibrary/media/Advisory_Circular/AC_90-120.pdf},
  note         = {Sections 2.12.1.4–2.12.1.5, \emph{ACAS Failures and Anomalies}}
}

@article{smith2020view,
  title={A view from the cockpit: Exploring pilot reactions to attacks on avionic systems},
  author={Smith, Matthew and Strohmeier, Martin and Harman, Jon and Lenders, Vincent and Martinovic, Ivan},
  year={2020},
  publisher={Internet Society}
}

@article{ribaric2023aviation,
  title        = {Aviation Cyber Security},
  author       = {Ribari{\'c}, Boris Z. and Vasiljevi{\'c}, Dragan and Vasiljevi{\'c}, Julijana and Mikanovi{\'c}, Boris R.},
  journal      = {Transport and Traffic Theory and Practice (TTTP)},
  year         = {2023},
  doi          = {10.7251/JTTTP2302037R},
  note         = {Received: September 13, 2023; Accepted: September 29, 2023},
}

@article{geister2018impact,
  title={Impact study on cyber threats to GNSS and FMS systems},
  author={Geister, Robert Manuel and Buch, Jan-Philipp and Niedermeier, Dominik and Gamba, Giovanni and Canzian, Luca and Pozzobon, Oscar},
  year={2018}
}

@inproceedings{truffer2017jamming,
  title={Jamming of aviation GPS receivers: Investigation of field trials performed with civil and military aircraft},
  author={Truffer, Pascal and Scaramuzza, Maurizio and Troller, Marc and Bertschi, Marc},
  booktitle={Proceedings of the 30th International Technical Meeting of the Satellite Division of The Institute of Navigation (ION GNSS+ 2017)},
  pages={1258--1266},
  year={2017}
}

@article{liu2024optimal,
  title={Optimal attack path planning based on reinforcement learning and cyber threat knowledge graph combining the ATT\&CK for air traffic management system},
  author={Liu, Chao and Wang, Buhong and Li, Fan and Tian, Jiwei and Yang, Yong and Luo, Peng and Liu, Zhouzhou},
  journal={IEEE Transactions on Transportation Electrification},
  year={2024},
  publisher={IEEE}
}

@book{howard2006security,
  title={The security development lifecycle},
  author={Howard, Michael and Lipner, Steve},
  volume={8},
  year={2006},
  publisher={Microsoft Press Redmond}
}

@INPROCEEDINGS{8260283,
  author={Khan, Rafiullah and McLaughlin, Kieran and Laverty, David and Sezer, Sakir},
  booktitle={2017 IEEE PES Innovative Smart Grid Technologies Conference Europe (ISGT-Europe)}, 
  title={STRIDE-based threat modeling for cyber-physical systems}, 
  year={2017},
  volume={},
  number={},
  pages={1-6},
  doi={10.1109/ISGTEurope.2017.8260283}}

@article{strom2018mitre,
  title={Mitre ATT\&CK: Design and philosophy: Technical report},
  author={Strom, Blake E and Applebaum, Andy and Miller, Doug P and Nickels, Kathryn C and Pennington, Adam G and Thomas, Cody B},
  journal={He MITRE Corp},
  year={2018}
}

@article{no2025study,
  title={A Study on Flight Crew’s Resilient Behavior Through Integration of Safety-I and Safety-II: Analysis of Aviation Safety Cases},
  author={No, Hyun Woo and Cha, Woo Chang},
  year={2025},
  publisher={Preprints}
}

@article{marinos2020aviation,
  title={AVIATION CYBERSECURITY: FAA Should Fully Implement Key Practices to Strengthen Its Oversight of Avionics Risks},
  author={Marinos, Nick},
  year={2020}
}

@article{wolf2014information,
  title={Information technology security threats to modern e-enabled aircraft: A cautionary note},
  author={Wolf, Marko and Minzlaff, Moritz and Moser, Martin},
  journal={Journal of Aerospace Information Systems},
  volume={11},
  number={7},
  pages={447--457},
  year={2014},
  publisher={American Institute of Aeronautics and Astronautics}
}

@article{ukwandu2022cyber,
  title={Cyber-security challenges in aviation industry: A review of current and future trends},
  author={Ukwandu, Elochukwu and Ben-Farah, Mohamed Amine and Hindy, Hanan and Bures, Miroslav and Atkinson, Robert and Tachtatzis, Christos and Andonovic, Ivan and Bellekens, Xavier},
  journal={Information},
  volume={13},
  number={3},
  pages={146},
  year={2022},
  publisher={MDPI}
}

@inproceedings{melnichuk2019development,
  title={Development of electronic flight bag software based on expert system for computing of optimal aircraft performance},
  author={Melnichuk, Alexander and Nesterov, Victor and Sudakov, Vladimir and Kirill, Sypalo},
  booktitle={2019 Twelfth International Conference" Management of large-scale system development"(MLSD)},
  pages={1--4},
  year={2019},
  organization={IEEE}
}

@inproceedings{bhardwaj2019safety,
  title={Safety and human factors for electronic flight bag usage in general aviation},
  author={Bhardwaj, Pranay and Purdy, Carla},
  booktitle={2019 IEEE National Aerospace and Electronics Conference (NAECON)},
  pages={181--184},
  year={2019},
  organization={IEEE}
}

@article{samosir2021effect,
  title={Effect of Effectiveness of Use of Electronic Flight Bags on Flight Safety at PT. Garuda Indonesia},
  author={Samosir, Johar and Sihombing, Sarinah and Kuntohadi, Hendro and Kurniawan, Johannes and Akbar, Abdulrachman Naufal},
  journal={Annals of the Romanian Society for Cell Biology},
  volume={25},
  number={3},
  pages={112--122},
  year={2021},
  publisher={" Vasile Goldis" Western University Arad, Romania}
}

@article{mecham2002new,
  title={New 777 Introduces Electronic Flight Bag},
  author={Mecham, Michael},
  journal={Aviation Week \& Space Technology},
  volume={157},
  number={23},
  pages={64--64},
  year={2002},
  publisher={Mcgraw Hill}
}

@inproceedings{lupidi2015contributing,
  title={Contributing towards sustainable aviation through an electronic flight bag for processing signals from avionic polarimetric weather radars},
  author={Lupidi, Alberto and Lischi, Stefano and Berizzi, Fabrizio and Baldini, L and Facheris, L and Cuccoli, F and others},
  booktitle={2015 International Symposium on Sustainable Aviation (ISSA)},
  year={2015}
}

@inproceedings{zelazo2012electronic,
  title={An electronic flight bag for NextGen avionics},
  author={Zelazo, Eyton},
  booktitle={Head-and Helmet-Mounted Displays XVII; and Display Technologies and Applications for Defense, Security, and Avionics VI},
  volume={8383},
  pages={171--176},
  year={2012},
  organization={SPIE}
}

@techreport{chandra2003human,
  title={Human factors considerations in the design and evaluation of electronic flight bags (EFBs): Version 2},
  author={Chandra, Divya C and Yeh, Michelle and Riley, Vic and Mangold, Susan J and others},
  year={2003},
  institution={United States. Department of Transportation. Federal Aviation Administration}
}

@mastersthesis{suppiah2019impact,
  title={Impact of electronic flight bag on pilot workload},
  author={Suppiah, Saravanan},
  school={Embry-Riddle Aeronautical University},
  address={Daytona Beach, Florida},
  year={2019},
}

@inproceedings{lopes2022supporting,
  title={Supporting situational awareness on aviation pilots: key insights affecting the use of electronic flight bags devices},
  author={Lopes, Nuno Moura and Aparicio, Manuela and Neves, F{\'a}tima Trindade},
  booktitle={World Conference on Information Systems and Technologies},
  pages={93--101},
  year={2022},
  organization={Springer}
}

@article{HowSecureAreIFECSystems2017,
  title        = {How Secure Are IFEC Systems?},
  author       = {Anonymous},
  month        = {Dec},
  year         = {2017},
  url          = {https://interactive.aviationtoday.com/how-secure-are-ifec-systems/},
  note         = {Retrieved from Aviation Today website}
}

@inproceedings{shetty2008system,
  title={System of systems design for worldwide commercial aircraft networks},
  author={Shetty, Sudhakar},
  booktitle={at 26th International Congress of the Aeronautical Sciences},
  year={2008}
}

@misc{pentestpartners2023efb1,
  author       = {Pen Test Partners},
  title        = {EFB Tampering Part 1: Introduction and Class Differences},
  year         = {2023},
  url          = {https://www.pentestpartners.com/security-blog/efb-tampering-1-introduction-and-class-differences/},
  note         = {Accessed: 2025-10-31}
}

@misc{pentestpartners2023efb2,
  author       = {Pen Test Partners},
  title        = {EFB Tampering Part 2: Device Integrity},
  year         = {2023},
  url          = {https://www.pentestpartners.com/security-blog/efb-tampering-2-device-integrity/},
  note         = {Accessed: 2025-10-31}
}

@misc{pentestpartners2023efb3a,
  author       = {Pen Test Partners},
  title        = {EFB Tampering Part 3: Take-Off (Pt.1)},
  year         = {2023},
  url          = {https://www.pentestpartners.com/security-blog/efb-tampering-3-take-off-pt1/},
  note         = {Accessed: 2025-10-31}
}

@misc{pentestpartners2023efb3b,
  author       = {Pen Test Partners},
  title        = {EFB Tampering Part 3: Take-Off (Pt.2)},
  year         = {2023},
  url          = {https://www.pentestpartners.com/security-blog/efb-tampering-3-take-off-pt2/},
  note         = {Accessed: 2025-10-31}
}

@article{bitton2019machine,
  title={A machine learning-based intrusion detection system for securing remote desktop connections to electronic flight bag servers},
  author={Bitton, Ron and Shabtai, Asaf},
  journal={IEEE Transactions on Dependable and Secure Computing},
  volume={18},
  number={3},
  pages={1164--1181},
  year={2019},
  publisher={IEEE}
}

@inproceedings{true2021cybersecurity,
  title={Cybersecurity for flight deck data exchange},
  author={True, Willard and Kilbourne, Todd and Roy, Aloke and Ghazavi, Noureddin},
  booktitle={2021 IEEE/AIAA 40th Digital Avionics Systems Conference (DASC)},
  pages={1--13},
  year={2021},
  organization={IEEE}
}

@inproceedings{lewis2025developing,
  title={Developing a Cybersecurity Architecture for Extensible Traffic Management (xTM)},
  author={Lewis, Terrence and Ali, Hassan and Freeman, Kenneth},
  booktitle={AIAA SCITECH 2025 Forum},
  pages={2720},
  year={2025}
}

@inproceedings{freeman2022immutable,
  title={Immutable secure data exchange and storage for urban air mobility environments},
  author={Freeman, Kenneth and Garcia, Steve W},
  booktitle={AIAA SCITECH 2022 Forum},
  pages={1092},
  year={2022}
}

@misc{Samanani_2025, title={Implementing comprehensive cybersecurity to secure the Advanced Air Mobility Ecosystem}, url={https://www.skygrid.com/implementing-comprehensive-cybersecurity-to-secure-the-advanced-air-mobility-ecosystem/}, journal={SkyGrid}, publisher={SkyGrid}, author={Samanani, Shezan}, year={2025}, month={May}}

@article{saleem2025cybersecurity,
  title={Cybersecurity Risks in EV Mobile Applications: A Comparative Assessment of OEM and Third-Party Solutions},
  author={Saleem, Bilal and Rehman, Alishba and Hassan, Muhammad Ali and Muhammad, Zia},
  journal={World Electric Vehicle Journal},
  volume={16},
  number={7},
  pages={364},
  year={2025},
  publisher={MDPI}
}

@inproceedings{xiang2022multi,
  title={A Multi-stage Precision Landing Method for Autonomous eVTOL Based on Multi-marker Joint Localization},
  author={Xiang, Senwei and Ye, Minxiang and Zhu, Shiqiang and Gu, Jason and Xie, Anhuan and Men, Zehua},
  booktitle={2022 IEEE International Conference on Robotics and Biomimetics (ROBIO)},
  pages={1--6},
  year={2022},
  organization={IEEE}
}

@article{xiang2024autonomous,
  title={Autonomous eVTOL: A summary of researches and challenges},
  author={Xiang, Senwei and Xie, Anhuan and Ye, Minxiang and Yan, Xufei and Han, Xiaojia and Niu, Hongjiao and Li, Qiang and Huang, Haishan},
  journal={Green Energy and Intelligent Transportation},
  volume={3},
  number={1},
  pages={100140},
  year={2024},
  publisher={Elsevier}
}

@inproceedings{brown2024visual,
  title={Visual \& inertial datasets for an eVTOL aircraft approach and landing scenario},
  author={Brown, Nelson and Kawamura, Evan and Bard, Luke and Jaffe, AJ and Ringelberg, Wayne and Kannan, Keerthana and Ippolito, Corey A},
  booktitle={AIAA SciTech 2024 Forum},
  pages={1386},
  year={2024}
}

@inproceedings{mendonca2022advanced,
  title={Advanced air mobility vertiport considerations: A list and overview},
  author={Mendonca, Nancy and Murphy, James and Patterson, Michael D and Alexander, Rex and Juarex, Gabriela and Harper, Clint},
  booktitle={AIAA AVIATION 2022 Forum},
  pages={4073},
  year={2022}
}

@article{veneruso2022sensing,
  title={Sensing requirements and vision-aided navigation algorithms for vertical landing in good and low visibility UAM scenarios},
  author={Veneruso, Paolo and Opromolla, Roberto and Tiana, Carlo and Gentile, Giacomo and Fasano, Giancarmine},
  journal={Remote Sensing},
  volume={14},
  number={15},
  pages={3764},
  year={2022},
  publisher={MDPI}
}

@article{brown2017adversarial,
  title={Adversarial patch},
  author={Brown, Tom B and Man{\'e}, Dandelion and Roy, Aurko and Abadi, Mart{\'\i}n and Gilmer, Justin},
  journal={arXiv preprint arXiv:1712.09665},
  year={2017}
}

@inproceedings{cao2021invisible,
  title={Invisible for both camera and LiDAR: Security of multi-sensor fusion based perception in autonomous driving under physical-world attacks},
  author={Cao, Yulong and Wang, Ningfei and Xiao, Chaowei and Yang, Dawei and Fang, Jin and Yang, Ruigang and Chen, Qi Alfred and Liu, Mingyan and Li, Bo},
  booktitle={2021 IEEE symposium on security and privacy (SP)},
  pages={176--194},
  year={2021},
  organization={IEEE}
}

@article{hendrycks2019robustness,
  title={Benchmarking Neural Network Robustness to Common Corruptions and Perturbations},
  author={Dan Hendrycks and Thomas Dietterich},
  journal={Proceedings of the International Conference on Learning Representations},
  year={2019}
}

@article{lu2021scale,
  title={Scale-adaptive adversarial patch attack for remote sensing image aircraft detection},
  author={Lu, Mingming and Li, Qi and Chen, Li and Li, Haifeng},
  journal={Remote Sensing},
  volume={13},
  number={20},
  pages={4078},
  year={2021},
  publisher={MDPI}
}

@inproceedings{kawamura2025vision,
  title={Vision-Based Distributed Sensing at Vertiports for Advanced Air Mobility and Urban Air Mobility Approach and Landing},
  author={Kawamura, Evan and Kannan, Keerthana and Lombaerts, Thomas and Stepanyan, Vahram and Dolph, Chester and Brown, Nelson and Ippolito, Corey A},
  booktitle={AIAA SCITECH 2025 Forum},
  pages={0346},
  year={2025}
}

@inproceedings{ippolito2025enabling,
  title={Enabling Smart Urban Airspaces through Distributed Sensing Technologies},
  author={Ippolito, Corey A and Martin, Rodney A and Kawamura, Evan and Gorospe, George and Holforty, Wendy and Kannan, Keerthana and Stepanyan, Vahram and Lombaerts, Thomas and Brown, Nelson and Dolph, Chester},
  booktitle={AIAA SCITECH 2025 Forum},
  pages={0344},
  year={2025}
}

@article{hu2025development,
  title={Development and challenges of autonomous electric vertical take-off and landing aircraft},
  author={Hu, Lijuan and Yan, Xufei and Yuan, Ye},
  journal={Heliyon},
  volume={11},
  number={1},
  year={2025},
  publisher={Elsevier}
}

@article{hendrycks2020augmix,
  title={{AugMix}: A Simple Data Processing Method to Improve Robustness and Uncertainty},
  author={Hendrycks, Dan and Mu, Norman and Cubuk, Ekin D. and Zoph, Barret and Gilmer, Justin and Lakshminarayanan, Balaji},
  journal={Proceedings of the International Conference on Learning Representations (ICLR)},
  year={2020}
}

@article{hendrycks2019oe,
  title={Deep Anomaly Detection with Outlier Exposure},
  author={Hendrycks, Dan and Mazeika, Mantas and Dietterich, Thomas},
  journal={Proceedings of the International Conference on Learning Representations},
  year={2019}
}

@article{liu2020energy,
      title={Energy-based Out-of-distribution Detection},
      author={Liu, Weitang and Wang, Xiaoyun and Owens, John and Li, Yixuan},
      journal={Advances in Neural Information Processing Systems},
      year={2020}
 }

@inproceedings{xiang2021patchguard,
  title={$\{$PatchGuard$\}$: A provably robust defense against adversarial patches via small receptive fields and masking},
  author={Xiang, Chong and Bhagoji, Arjun Nitin and Sehwag, Vikash and Mittal, Prateek},
  booktitle={30th USENIX Security Symposium (USENIX Security 21)},
  pages={2237--2254},
  year={2021}
}

@inproceedings{xiang2022patchcleanser,
  title={$\{$PatchCleanser$\}$: Certifiably robust defense against adversarial patches for any image classifier},
  author={Xiang, Chong and Mahloujifar, Saeed and Mittal, Prateek},
  booktitle={31st USENIX security symposium (USENIX Security 22)},
  pages={2065--2082},
  year={2022}
}

@inproceedings{wang2019neural,
  title={Neural cleanse: Identifying and mitigating backdoor attacks in neural networks},
  author={Wang, Bolun and Yao, Yuanshun and Shan, Shawn and Li, Huiying and Viswanath, Bimal and Zheng, Haitao and Zhao, Ben Y},
  booktitle={2019 IEEE symposium on security and privacy (SP)},
  pages={707--723},
  year={2019},
  organization={IEEE}
}

@inproceedings{gao2019strip,
  title={Strip: A defence against trojan attacks on deep neural networks},
  author={Gao, Yansong and Xu, Change and Wang, Derui and Chen, Shiping and Ranasinghe, Damith C and Nepal, Surya},
  booktitle={Proceedings of the 35th annual computer security applications conference},
  pages={113--125},
  year={2019}
}

@article{levitt2023uam,
  title={Uam airspace research roadmap-rev. 2.0},
  author={Levitt, Ian and Phojanamongkolkij, Nipa and Horn, Adam and Witzberger, Kevin},
  year={2023}
}

@article{zhang2023cooperative,
  title={Cooperative perception for safe control of autonomous vehicles under lidar spoofing attacks},
  author={Zhang, Hongchao and Li, Zhouchi and Cheng, Shiyu and Clark, Andrew},
  journal={arXiv preprint arXiv:2302.07341},
  year={2023}
}

@article{mishra2023autonomous,
  title={Autonomous advanced aerial mobility—An end-to-end autonomy framework for UAVs and beyond},
  author={Mishra, Sakshi and Palanisamy, Praveen},
  journal={IEEE Access},
  volume={11},
  pages={136318--136349},
  year={2023},
  publisher={IEEE}
}

@article{pearce2021signal,
  author  = {Pearce, Ben and others},
  title   = {Signal injection attacks on ADS-B: Analysis and detection},
  journal = {IEEE Transactions on Aerospace and Electronic Systems},
  volume  = {57},
  number  = {5},
  pages   = {3451--3464},
  year    = {2021},
  publisher = {IEEE}
}

@article{slimane2022ads,
  author  = {Slimane, Hafedh and others},
  title   = {ADS-B vulnerabilities and countermeasures: A comprehensive survey},
  journal = {Journal of Air Transport Management},
  volume  = {104},
  pages   = {102203},
  year    = {2022},
  publisher = {Elsevier}
}

@inproceedings{shang2019multidevice,
  author    = {Shang, Bin and Zhao, Xinyue and Wang, Jian},
  title     = {Multi-device spoofing attacks on ADS-B systems using SDRs},
  booktitle = {IEEE/AIAA Digital Avionics Systems Conference (DASC)},
  year      = {2019},
  pages     = {1--10},
  publisher = {IEEE}
}

@article{khan2024survey,
  author  = {Khan, Md. and others},
  title   = {A survey on ADS-B spoofing and detection methods for modern aviation},
  journal = {Aerospace Science and Technology},
  volume  = {150},
  pages   = {108216},
  year    = {2024},
  publisher = {Elsevier}
}

@inproceedings{popper2011investigation,
  author    = {Popper, Christoph and Capkun, Srdjan},
  title     = {Investigation of signal manipulation attacks on ADS-B systems},
  booktitle = {Proceedings of the ACM Conference on Wireless Network Security (WiSec)},
  year      = {2011},
  pages     = {331--340},
  publisher = {ACM}
}

@techreport{tang2021uamcyber,
  author      = {Tang, Yulong and others},
  title       = {Cybersecurity Considerations for Urban Air Mobility},
  institution = {NASA Ames Research Center},
  year        = {2021},
  note        = {NASA Technical Report, UAM Cybersecurity Assessment}
}

@techreport{stouffer2020cns,
  author      = {Stouffer, Keith and others},
  title       = {Communications, Navigation, and Surveillance (CNS) Architecture for Urban Air Mobility},
  institution = {National Institute of Standards and Technology (NIST)},
  year        = {2020},
  note        = {NIST Technical Report}
}

@techreport{faa2016uasC2,
  title       = {Command and Control (C2) Link for Unmanned Aircraft Systems (UAS)},
  institution = {Federal Aviation Administration (FAA)},
  year        = {2016},
  note        = {FAA Spectrum Designation 5030–5091 MHz for UAS C2}
}

@inproceedings{ullah2025cns5g,
  author    = {Ullah, Imran and Khan, Shams and Alsaadi, Faisal E.},
  title     = {Leveraging 5G Communication for CNS in Urban Air Mobility: Challenges and Security Implications},
  booktitle = {AIAA Aviation Forum},
  year      = {2025},
  organization = {AIAA}
}

@inproceedings{kuba2024tcas,
  author    = {Kuba, S. and Babiceanu, R. F.},
  title     = {Navigating Threats: A Vulnerability Analysis of TCAS Interaction with Other Aircraft Systems},
  booktitle = {AIAA DATC/IEEE 43rd Digital Avionics Systems Conference (DASC)},
  year      = {2024},
  pages     = {1--6},
  publisher = {IEEE}
}

@inproceedings{smith2020cockpit,
  author    = {Smith, M. and Strohmeier, M. and Harman, J. and Lenders, V. and Martinovic, I.},
  title     = {A View from the Cockpit: Exploring Pilot Reactions to Attacks on Avionic Systems},
  booktitle = {IEEE/AIAA Digital Avionics Systems Conference (DASC)},
  year      = {2020},
  pages     = {1--10},
  publisher = {IEEE}
}

@techreport{faa2024acas,
  author      = {{Federal Aviation Administration}},
  title       = {Advisory Circular AC 90-120: Operational Use of Airborne Collision Avoidance Systems},
  institution = {U.S. Department of Transportation, Federal Aviation Administration},
  year        = {2024},
  url         = {https://www.faa.gov/documentLibrary/media/Advisory_Circular/AC_90-120.pdf},
  note        = {Sections 2.12.1.4–2.12.1.5, ACAS Failures and Anomalies}
}

@incollection{strohmeier2020airground,
  author    = {Strohmeier, M. and Martinovic, I. and Lenders, V.},
  title     = {Securing the Air–Ground Link in Aviation},
  booktitle = {The Security of Critical Infrastructures: Risk, Resilience and Defense},
  editor    = {B{\'o}na, G{\'a}bor},
  pages     = {131--154},
  publisher = {Springer},
  year      = {2020}
}

@article{ukwandu2022cyberaviation,
  author  = {Ukwandu, E. and Ben-Farah, M. A. and Hindy, H. and Bures, M. and Atkinson, R. and Tachtatzis, C. and Andonovic, I. and Bellekens, X.},
  title   = {Cyber-Security Challenges in the Aviation Industry: A Review of Current and Future Trends},
  journal = {Information},
  volume  = {13},
  number  = {3},
  pages   = {146},
  year    = {2022},
  publisher = {MDPI}
}

@inproceedings{fonyi2024_5gsec,
  author    = {Fonyi, Bence and T{\"o}r{\"o}k, P{\'e}ter and Kov{\'a}cs, D{\'a}niel},
  title     = {Security of 5G-Enabled Communications for Aerial Mobility and Cooperative UAV Networks},
  booktitle = {IEEE Aerospace Conference},
  year      = {2024},
  pages     = {1--10},
  publisher = {IEEE}
}

@techreport{mitre2023uam5g,
  author      = {The MITRE Corporation},
  title       = {Security Analysis of 5G Connectivity for Urban Air Mobility Systems},
  institution = {MITRE Technical Report},
  year        = {2023}
}

@article{campbell2020oauth,
  title={OAuth 2.0 mutual-TLS client authentication and certificate-bound access tokens},
  author={Campbell, Brian and Bradley, John and Sakimura, Nat and Lodderstedt, Torsten},
  journal={Internet Requests for Comments, IETF, RFC 8705},
  year={2020}
}

@techreport{rescorla2018transport,
  title={The transport layer security (TLS) protocol version 1.3},
  author={Rescorla, Eric},
  year={2018}
}

@article{zhong2025enhancing,
  title={Enhancing aircraft reliability with information redundancy: A sensor-modal fusion approach leveraging deep learning},
  author={Zhong, Jie and Zhang, Heng and Miao, Qiang},
  journal={Reliability Engineering \& System Safety},
  volume={261},
  pages={111068},
  year={2025},
  publisher={Elsevier}
}

@article{voice-cloning,
    author={Barrington, Sarah and Cooper, Emily A. and Farid, Hany},
    title={People are poorly equipped to detect AI-powered voice clones},
    journal={Scientific Reports},
    year={2025},
    month={Mar},
    day={31},
    volume={15},
    number={1},
    pages={11004},
    issn={2045-2322},
    doi={10.1038/s41598-025-94170-3},
    url={https://doi.org/10.1038/s41598-025-94170-3},
}

@article{rfi-aviation,
    AUTHOR = {Malik, Adnan and Rao, Muzaffar},
    TITLE = {Radio Frequency Interference, Its Mitigation and Its Implications for the Civil Aviation Industry},
    JOURNAL = {Electronics},
    VOLUME = {14},
    YEAR = {2025},
    NUMBER = {12},
    ARTICLE-NUMBER = {2483},
    URL = {https://www.mdpi.com/2079-9292/14/12/2483},
    ISSN = {2079-9292},
    DOI = {10.3390/electronics14122483},
}

@techreport{AAM-implementation-plan,
  title        = {Advanced Air Mobility Implementation Plan},
  author       = {{Federal Aviation Administration}},
  institution  = {U.S. Department of Transportation, Federal Aviation Administration},
  year         = {2023},
  month        = {July},
  number       = {Version 1.0},
  note         = {Available online: \url{https://www.faa.gov/sites/faa.gov/files/AAM-I28-Implementation-Plan.pdf}},
}

@techreport{eurocae-info-sec-vtol,
  title        = {ED-305 | Information Security Guidance for VTOL and Collaborative Systems},
  author       = {WG-112},
  institution  = {EUROCAE},
  year         = {2025},
  note         = {Available online: \url{https://www.eurocae.net/product/ed-305-information-security-guidance-for-vtol-and-collaborative-systems/}},
}

@article{rf-fp-vs-ia,
    title = {Analysis of impersonation attacks on systems using RF fingerprinting and low-end receivers},
    journal = {Journal of Computer and System Sciences},
    volume = {80},
    number = {3},
    pages = {591-601},
    year = {2014},
    note = {Special Issue on Wireless Network Intrusion},
    issn = {0022-0000},
    doi = {https://doi.org/10.1016/j.jcss.2013.06.013},
    url = {https://www.sciencedirect.com/science/article/pii/S0022000013001220},
    author = {Saeed Ur Rehman and Kevin W. Sowerby and Colin Coghill},
}

@article{rf-fp-review,
  author={Soltanieh, Naeimeh and Norouzi, Yaser and Yang, Yang and Karmakar, Nemai Chandra},
  journal={IEEE Journal of Radio Frequency Identification}, 
  title={A Review of Radio Frequency Fingerprinting Techniques}, 
  year={2020},
  volume={4},
  number={3},
  pages={222-233},
  doi={10.1109/JRFID.2020.2968369},
}

@misc{volocopter2023demoflight,
  author       = {Volocopter GmbH},
  title        = {Volocopter Demonstration Flight at Groupe ADP / Skyports Vertiport Testbed},
  year         = {2023},
  howpublished = {\url{https://www.volocopter.com}},
  note         = {Accessed: 2025-11-16}
}

@misc{skydrive2024vertiport,
  author       = {SkyDrive, Inc.},
  title        = {SkyDrive SD-05 eVTOL at Urban Vertiport Demonstration},
  year         = {2024},
  howpublished = {\url{https://en.skydrive2020.com}},
  note         = {Accessed: 2025-11-16}
}

@article{trask2024arinc,
  title={ARINC 429 Cyber-vulnerabilities and Voltage Data in a Hardware-in-the-Loop Simulator},
  author={Trask, Connor and Movit, Steve and Clutter, Justace and Clark, Rosene and Herrera, Mark and Tran, Kelly},
  journal={arXiv preprint arXiv:2408.16714},
  year={2024}
}

@techreport{Rieder2023AAMSecurity,
  title        = {Security Considerations for Advanced Air Mobility (AAM) Operations at Airports},
  author       = {René Rieder, Jr.},
  institution  = {Burns Engineering, Inc.},
  type         = {PARAS Technical Report},
  number       = {PARAS 0041},
  year         = {2023},
  month        = {June},
  url          = {https://www.sskies.org/images/uploads/subpage/PARAS_0041.AAMOperations_.FinalReport_.pdf},
  note         = {Program for Applied Research in Airport Security, National Safe Skies Alliance}
}

@misc{NavblueNavigationPlus,
  title        = {Navigation+ },
  howpublished          = {https://www.navblue.aero/product/navigation-plus/},
  year ={2025},
  note      = {Accessed: 2025-11-21}
}

@misc{JeppesenNavData,
  title        = {NavData -- Aeronautical Navigation Data},
  howpublished = {\url{https://ww2.jeppesen.com/navigation-solutions/navdata/}},
  year         = {2025},
  note         = {Accessed: 2025-11-21}
}

@misc{LidoDeveloperPortal2024,
  author       = {Lufthansa Systems},
  title        = {Lido Developer Portal: Easily accessible, high-quality aeronautical data},
  url          = {https://cdn.lhsystems.com/2024-05/Lido_Developer_Portal.pdf},
  note         = {Accessed: 2025-11-21},
  year         = {2024}
}

@misc{USP20200013243A1,
  title        = {Data Corruption Detection in Navigation Databases},
  author       = {Anonymous},
  year         = {2020},
  number       = {US20200013243A1},
  url          = {https://patents.google.com/patent/US20200013243A1/en},
  note         = {Describes detection and mitigation of tampering in flight management system navigation databases}
}

@article{Martinez2021,
  title        = {Software Supply Chain Attacks — A Threat to Global Cybersecurity: SolarWinds’ Case Study},
  author       = {Javier Martinez and Alejandro Duran},
  journal      = {International Journal of Security and Its Applications},
  volume       = {15},
  number       = {5},
  pages        = {123--136},
  year         = {2021},
  doi          = {10.18280/ijsse.110505}
}

@article{Dong2025,
  title        = {The '4W+1H' of Software Supply Chain Security Checklist for Critical Infrastructure},
  author       = {Xin Dong and Hyun Lee and Wei Xing and Muhammad Ahmed and Georgios Avgoustakis},
  journal      = {CoRR},
  volume       = {abs/2510.26174},
  year         = {2025},
  note         = {arXiv:2510.26174}
}

@inproceedings{Okafor2024,
  title        = {SoK: Analysis of Software Supply Chain Security by Establishing Secure Design Properties},
  author       = {Chinenye Okafor and Taylor R. Schorlemmer and Santiago Torres-Arias and James C. Davis},
  booktitle    = {Proceedings of the 1st ACM Workshop on Software Supply Chain Offensive Research and Ecosystem Defenses (SCORED)},
  pages        = {15--24},
  year         = {2024},
  doi          = {10.1145/3560835.3564556}
}

@article{Welsh2025,
  title        = {Towards Socio-Technical Topology-Aware Adaptive Threat Detection in Software Supply Chains},
  author       = {Brendan Welsh and Jón Finnsson and Stefán Stefánsson and Klaus Neukirchen},
  journal      = {CoRR},
  volume       = {abs/2510.21452},
  year         = {2025},
  note         = {arXiv:2510.21452}
}

@article{altaweel2023gps,
  title={GPS spoofing attacks in fanets: A systematic literature review},
  author={Altaweel, Ala and Mukkath, Hena and Kamel, Ibrahim},
  journal={IEEE Access},
  volume={11},
  pages={55233--55280},
  year={2023},
  publisher={IEEE}
}

@article{jafarnia2012gps,
  title={GPS vulnerability to spoofing threats and a review of antispoofing techniques},
  author={Jafarnia-Jahromi, Ali and Broumandan, Ali and Nielsen, John and Lachapelle, G{\'e}rard},
  journal={International Journal of Navigation and Observation},
  volume={2012},
  number={1},
  pages={127072},
  year={2012},
  publisher={Wiley Online Library}
}

@inproceedings{hu2009study,
  title={A study of GPS jamming and anti-jamming},
  author={Hu, Hui and Wei, Na},
  booktitle={2009 2nd international conference on power electronics and intelligent transportation system (PEITS)},
  volume={1},
  pages={388--391},
  year={2009},
  organization={IEEE}
}

@article{li2023overview,
  title={Overview of jamming technology for satellite navigation},
  author={Li, Xiangjun and Chen, Lei and Lu, Zukun and Wang, Feixue and Liu, Wenxiang and Xiao, Wei and Liu, Peiguo},
  journal={Machines},
  volume={11},
  number={7},
  pages={768},
  year={2023},
  publisher={MDPI}
}

@misc{cisa2025tcas,
  title        = {CISA Advisory ICSA-25-021-01: Traffic Collision Avoidance System (TCAS II)},
  author       = {{Cybersecurity and Infrastructure Security Agency}},
  year         = {2025},
  url          = {https://www.cisa.gov/news-events/ics-advisories/icsa-25-021-01?utm_source=chatgpt.com},
  note         = {Accessed: 2025-02-02}
}

@misc{foreflight2025mobile,
  title        = {ForeFlight Mobile Electronic Flight Bag},
  author       = {{ForeFlight LLC}},
  year         = {2025},
  note         = {Screenshots retrieved from the ForeFlight Mobile product page},
  url          = {https://foreflight.com/products/foreflight-mobile/},
  note         = {Accessed: 2025-02-02}
}

@ARTICLE{8255823,
  author={Shamaei, Kimia and Khalife, Joe and Kassas, Zaher M.},
  journal={IEEE Transactions on Wireless Communications}, 
  title={Exploiting LTE Signals for Navigation: Theory to Implementation}, 
  year={2018},
  volume={17},
  number={4},
  pages={2173-2189},
  doi={10.1109/TWC.2018.2789882}}

@ARTICLE{9541006,
  author={Khalife, Joe and Neinavaie, Mohammad and Kassas, Zaher M.},
  journal={IEEE Transactions on Aerospace and Electronic Systems}, 
  title={The First Carrier Phase Tracking and Positioning Results With Starlink LEO Satellite Signals}, 
  year={2022},
  volume={58},
  number={2},
  pages={1487-1491},
  doi={10.1109/TAES.2021.3113880}}

\end{document}